\documentclass[pra,amsmath,amssymb,amsfonts,twocolumn,
   superscriptaddress,nobalancelastpage,floatfix, nofootinbib]{revtex4}

\usepackage{bbm}
\usepackage{amssymb}
\usepackage{amsmath}
\usepackage{verbatim}
\usepackage[pdftex]{graphicx}
\usepackage{epsfig}
\usepackage{dsfont}
\usepackage{color}

   {\hspace*{\fill}$\Box$\vspace{1.5ex}\par}

\newcommand{\ad}{a^{\dagger}}

\newcommand{\tr}{\mbox{tr}}

\newcommand{\vac}{|0\rangle }

\begin{document}
\title{Generalized Hartree-Fock Theory for Interacting Fermions in Lattices: Numerical Methods}
\author{Christina V.\ Kraus}
\author{J.\ Ignacio Cirac}
\affiliation{Max-Planck-Institute for Quantum Optics, Hans-Kopfermann-Str.\ 1, D-85748 Garching,
Germany.}

\begin{abstract}
We present numerical methods to solve the Generalized Hartree-Fock theory 
for fermionic systems in lattices, both in thermal equilibrium and out of equilibrium. Specifically, we show how to determine the covariance matrix corresponding to the Fermionic Gaussian state that optimally approximates the quantum state of the fermions. The methods apply to relatively large systems, since their complexity only scales quadratically with the number of lattice sites. Moreover, they are specially suited to describe inhomogenous systems, as those typically found in recent experiments with atoms in optical lattices, at least in the weak interaction regime. As a benchmark, we have applied them to the two-dimensional Hubbard model on a $10 \times 10$ lattice with and without an external confinement.
\end{abstract}

\maketitle

\section{Introduction}

Recent experiments with cold atomic systems have attracted the interest of
several scientific communities. Many physical phenomena, like Bose-Einstein condensation, superfluidity, or the formation of Cooper pairs, have been observed in seminal experiments and carefully analyzed and characterized ~\cite{Stringari_Bosons, Stringari_Fermions}. In the last years, the possibility of loading the atoms in optical lattices offers a new avenue to the observation of many other intriguing phenomena ~\cite{bloch-2008-80}.

There exists several theoretical tools in order to describe most of those phenomena. For the case of Bosons, the Gross--Pitaevskii equation provides us with the basic tool to analyze many static and dynamic properties for cold atomic gases ~\cite{Piataevskii}. Such an equation is valid whenever the occupation number of a
particular mode is large, and it is equivalent to a mean--field theory.
Typically, it can be solved numerically, offering very accurate predictions and explanations of all the observed phenomena in the weak interacting regime. Furthermore, it can also be used to describe lattice systems, at least as long as strong correlations are absent. In the case of Fermions, the basic tools at hand are Hartree--Fock and BCS theory ~\cite{fetter:walecka}. The first one assumes that the state of the particles can be approximated in terms of a Slater determinant, and thus it also correspond to a mean--field theory. The latter assumes a specific form of the wavefunction, whereby pairing between Fermions is allowed. For both approaches, a time--dependent version has been constructed \cite{barankov-2004-93, barankov-2006-73, Barankov-PhysRevA.73.033614, hastings-2008-BCS, Massel}.

Hartree--Fock and BCS theories can be unified in terms of a more general framework, the generalized Hartree--Fock Theory (gHFT) ~\cite{bach-1993}. The main idea is that both, the Slater--Determinant-- and BCS--states are members of a larger family of states, the so--called Fermionic Gaussian states (FGS). Therefore, one may try to find the state within this family which best approaches the real state of the system under study.
FGS are those states whose density operator can be expressed as an exponential of a quadratic function of canonical creation and annihilation operators. This property immediately implies that they fulfill Wick's theorem: that is, all correlation functions can be expressed in terms of the second moments of all creation and annihilation operators. Such moments are gathered in a matrix, the so--called covariance matrix (CM) which completely characterizes the state ~\cite{LinearOptics}. Thus, FGS can be very efficiently described, since the number of parameters only scales quadratically with the number of available fermionic modes and the expectation values of observables can be easily computed. For the case of lattices, the number of modes is automatically finite, and thus it should be
an ideal playground for such a gHFT. In fact, Bach, Lieb and Solovej \cite{bach-1993}
were the first who applied such a theory to a Hubbard model in 2D, and where able to solve the corresponding equations exactly for the homogeneous case, both for the ground state and in thermal equilibrium.

gHFT can be easily applied to any lattice system, homogeneous or not. The resulting equations are, however, difficult to solve in general. In this paper we derive such equations for the CM, and introduce different methods to solve them. First of all, we derive the equation for the real time evolution of the CM. Second, we consider the state within the FGS which minimizes the
total energy, and thus approximates the ground state. For that, we
derive the evolution equation (in imaginary time) for the CM in the presence of a lattice Hamiltonian. We show that under such an equation, the CM converges to the one which minimizes\footnote{Strictly speaking, the CM converges to the solution for which the energy is extremal.} the energy within the family of FGS. Thus, the evolution equation provides us with a practical way of determining the approximation to the ground state. Finally, we
derive the equation for the CM that minimizes the free energy, and apply a simple fixed-point iteration method to solve it.

Then we apply the methods to the spin-1/2 Hubbard model in a $10\times 10$ lattice. First, we benchmark them with exact solutions of the gHFT (for the homogeneous case) \cite{bach-1993}, finding an extraordinary agreement both for attractive and repulsive interactions. Second, in order to test the validity of gHFT itself, we compare the results for positive interactions, where strongly correlated effects are expected to be most pronounced, with the Monte Carlo results of Refs. ~\cite{Scarletter, trivedi-2009}. The results are obviously less precise than those obtained with such methods, other Monte Carlo approaches \cite{corney-2004-93}, or the recently developed algorithms based on PEPS and other tensor networks states \cite{kraus-2009, corboz-2009,  pineda-2009, corboz-2009-80, barthel-2009-80, Zhou-2010}. Although we obtain a good qualitative agreement with ~\cite{Scarletter, trivedi-2009} in many regimes, the FGS are, as expected, not able to capture all possible fermionic phases in the strong-correlation regimes, as they constitute a subclass of all possible fermionic states. For example, the finite temperature Mott phase cannot be approximated well as a Gaussian state, and is thus absent in the phase diagram.

At this point we would like to warn the reader that in this work we will mainly concentrate on the possibilities offered by the numerical approaches, and not focus on the physics of the Hubbard model. To be precise, our goal is a)  to formulate the standard gHFT for fermionic systems with a two-body interaction in the lattice using the language of covariance matrices, b) to derive numerical methods to determine the ground and thermal states as well as the time evolution of these systems within gHFT and c) to benchmark our techniques by applying them to the two-dimensional Hubbard model. 

This paper is organized as follows: In Sec.~\ref{sec:statement} we give a precise formulation of the
problem we aim to solve, and introduce the concepts necessary for the understanding of FGS and gHFT. This Section includes further a discussion why
FGS are supposed to capture well the physical properties of some of the most
relevant fermionic models. Next, we derive the methods to approximate the real time evolution, the
ground and thermal states of fermionic systems within gHFT in
Secs.~\ref{sec:real_time}--~\ref{sec:thermal_state}. Then, we benchmark our methods by applying them to the two-dimensional translationally invariant Hubbard model, both in the attractive and the repulsive regime, in Sec.~\ref{sec:numerics}. In addition, motivated by the current possibility of implementing the Hubbard model
in more than one dimension with the help of optical lattices, we also investigate the $2d$ Hubbard model with an external confinement in Sec.~\ref{sec:Hubbard_trap}, and close with a study of dynamical processes, both for the translationally invariant and the trapped system, in Sec.~\ref{sec:Dynamics}.

\section{Statement of the Problem}\label{sec:statement}
In this Section we give a statement of the problem, summarize the properties of fermionic Gaussian
states that are necessary for the understanding of our work (for more details see
e.g.~\cite{LinearOptics}) and argue why these states capture the properties of many relevant physical
models.

In our work we aim at understanding the physical properties of fermionic systems in a lattice, like for
instance
\begin{equation}\label{eq:H_twobody}
H = -\sum_{kl}t_{kl} \ad_{k}a_l + \sum_{klmn} u_{klmn}\ad_k\ad_la_ma_n,
\end{equation}
where the $a_k$ denote fermionic mode operators obeying canonical anti-commutation relations (CAR),
$\{a_k, \ad_l\}=\delta_{kl}$ and $t_{kl},  u_{klmn} \in \mathds{C}$. These models describe fermionic atoms in optical lattices ~\cite{Hofstetter}, and one of the most prominent ones, the Hubbard model  \cite{Hubbard_original},  is expected to describe high-Tc superconductivity ~\cite{Anderson_highTc}. We
approach the problem in the Majorana picture and define  $c_{k}= \ad_k+a_k$, $ c_{k+
M}=(-i)(\ad_k-a_k)$, where $M$ is the number of modes and $k=1, \ldots, M$. These operators satisfy the
CAR $\{c_k,c_l\}=2 \delta_{kl}$. The Hamiltonian $H$ in this picture is given by
\begin{equation}\label{eq:Hint}
H(T,U) = i\sum_{kl} T_{kl}c_k c_l + \sum_{klmn} U_{klmn}c_kc_lc_mc_n ,
\end{equation}
where $T_{kl},U_{klmn} \in \mathds{R}$. The CAR allow us to antisymmetrize $T$ and $U$ such that $T^T =
-T$ while $U$ is antisymmetric under the exchange of any of two adjacent indices.

Our goal is to derive ground and thermal states as well as the time-evolution of systems characterized
by the Hamiltonian given in Eq.~\eqref{eq:H_twobody} within the family of FGS.
These are states that can be represented as an exponential of a quadratic form in the Majorana
operators, and thus they are fully characterized by their second moments collected in the real and
anti-symmetric covariance matrix (CM) $\Gamma_{kl}=\langle \frac{i}{2}[c_k,c_l]\rangle$ from which all
higher correlations can be obtained via Wick's theorem (see e.g. ~\cite{LinearOptics}):
\begin{equation}\label{eq:Wick}
i^p\tr[\rho c_{j_1}\ldots c_{j_{2p}}]= \mathrm{Pf}(\Gamma|_{j_1\ldots j_{2p}}),
\end{equation}
where $1 \leq j_1 < \ldots < j_{2p} \leq 2M$ and $\Gamma_{j_1, \ldots, j_{2p}}$ is the corresponding
$2p \times 2p$ submatrix of $\Gamma$. $\mathrm{Pf}(\Gamma_{j_1, \ldots, j_{2p}})^2= \mbox{det}
(\Gamma_{j_1, \ldots, j_{2p}})$ is called the Pfaffian\footnote{The sign of the Pfaffian is determined by the condition that the the term $\Gamma_{j_1j_2}\Gamma_{j_3j_4}\ldots \Gamma_{j_{2p-1}j_{2p}}$ appears with positive sign.}. $\Gamma$ is the CM of a physical state iff
$i\Gamma - \mathds{1} \leq 0$, while pure states have to fulfill $\Gamma^2 = - \mathds{1}$. Every pure
FGS is the ground state of a quadratic Hamiltonian
\begin{equation}
H_Q = i\sum_{kl}h_{kl}c_kc_l,
\end{equation}
with real and antisymmetric Hamiltonian matrix $h$. All Gaussian states remain Gaussian under the time
evolution governed by a quadratic Hamiltonian, and the CM transforms according to $\Gamma(t) =
O(t)\Gamma(0)O(t)^T$, where $O(t) = e^{4ht}$ is an orthogonal transformation.

Let us now explain why FGS provide a very powerful technique for the description of fermionic many-body systems. First, note that all pure states within this family can be brought into a standard form $|\Psi\rangle = \prod_k (u_k + v_k \ad_k \ad_{-k})|0\rangle,$ where $|u_k|^2+|v_k|^2=1\; \forall k$ via a change of basis~\cite{BlochMessiah}. Thus, the pure Gaussian
states include the BCS-states introduced in the theory of superconductivity, as well as all Hartree Fock states. Second, FGS also include all mixed states
for which Wick's theorem~\eqref{eq:Wick} applies, e.g., thermal states of quadratic Hamiltonians. Thus, all together, we see
that FGS include a variety of states which form the basis of several many--body phenomena.

\section{Interacting Fermions in generalized Hartree Fock theory}\label{sec:methods}
In the following Section we derive the theoretical framework that will allow us later on to simulate
fermionic many-body problems, like the Hubbard model, for big system sizes. This is achieved by approximating the state of the fermions by a FGS, something which is implemented by means of Wick's theorem.

\subsection{Real Time Evolution}\label{sec:real_time}

\begin{figure}[t]
\begin{minipage}[t]{\columnwidth}
\centering
\includegraphics[width=0.95\columnwidth]{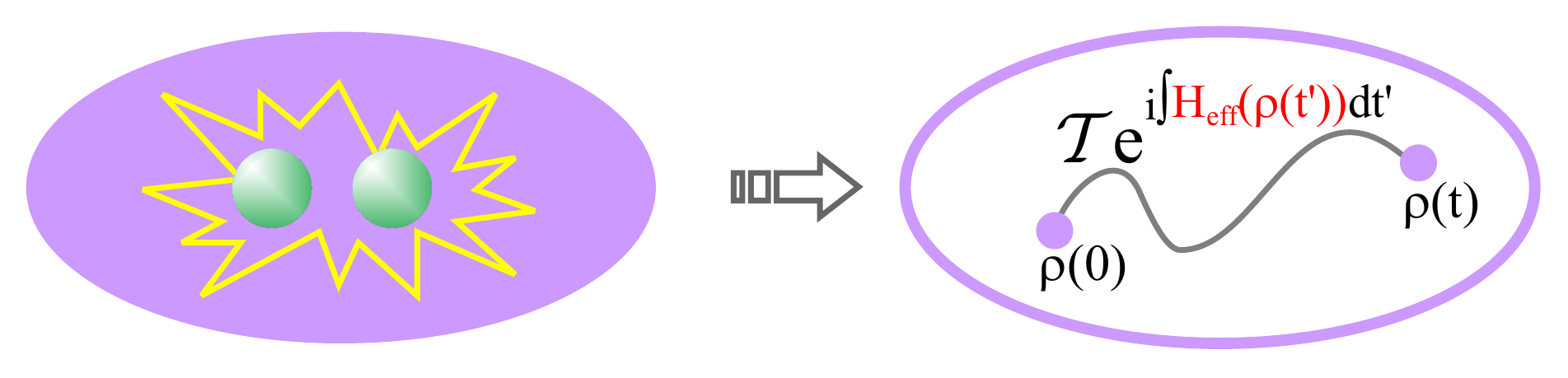}
\put(-250,50){ a)}
\end{minipage}
\begin{minipage}[t]{\columnwidth}
\centering
\includegraphics[width=0.95\columnwidth]{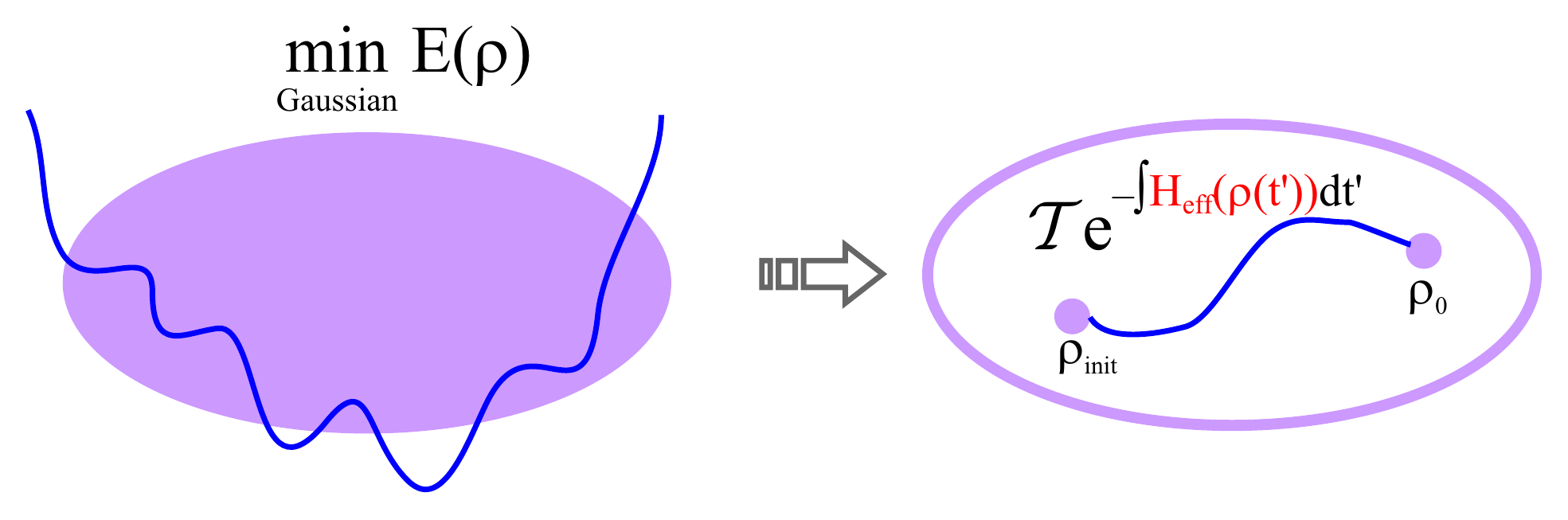}
\put(-250,70){ b)}
\end{minipage}
\begin{minipage}[t]{\columnwidth}
\centering
\includegraphics[width=0.95\columnwidth]{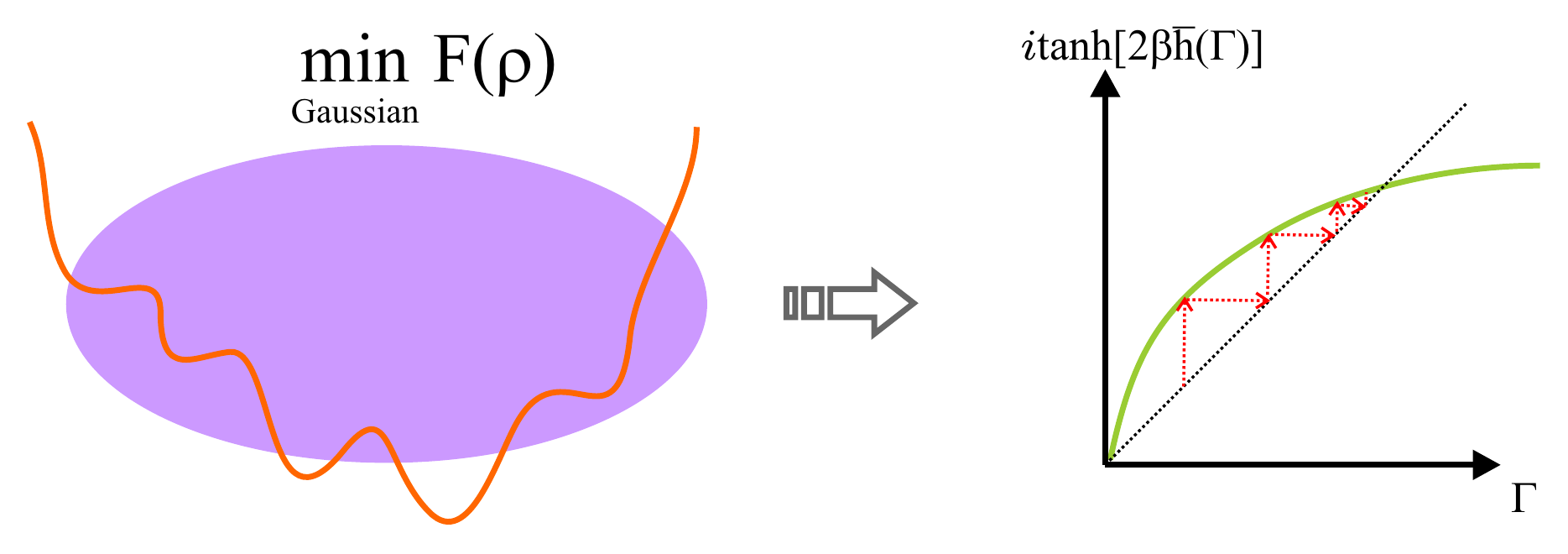}
\put(-250,70){ c)}
\end{minipage}
\caption{a) The time evolution under the interaction Hamiltonian \eqref{eq:H_twobody} within the set of
Gaussian states (purple) can be reformulated as the time-evolution under a \emph{quadratic but
state-dependent} Hamiltonian $H_{eff} = i\sum_{ij}\bar{h}_{ij}(\Gamma)c_ic_j$, where $\bar{h}$ is given
in Eq. \eqref{eq:hbar}, that respects the Gaussian structure. b) The minimization of the energy within the set of Gaussian states can be mapped to an imaginary time evolution within the set of Gaussian states with the \emph{effective, state-dependent} Hamiltonian
given by $H_{eff} = i\sum_{ij}\bar{h}_{ij}(\Gamma)c_ic_j$, with $\bar{h}$ defined in
Eq.~\eqref{eq:hbar}. c) The minimization of the free energy leads to the implicit equation $\Gamma = i \tanh
\left[2i\beta\bar{h}(\Gamma)\right]$ (c.f.~\eqref{eq:Gamma_Gibbs}) for the covariance matrix, that can
be solved, e.g., by a fixed-point iteration. \label{fig:Scheme}}
\end{figure}

We start the section deriving an evolution equation that allows us to describe the time-evolution of a
system at zero and finite temperature evolving under the interaction Hamiltonian~\eqref{eq:H_twobody}
within generalized Hartree-Fock theory. The
key idea is to use the covariance matrix for the description of the dynamical evolution. In the
Heisenberg picture, where the time evolution of the Majorana operators is given by $c_k(t) =
e^{iH(T,U)t}c_k e^{-iH(T,U)t}$, the covariance matrix evolves according to
\begin{equation}\label{eq:Gamma_dot_commutator}
\frac{d}{dt}\Gamma_{\alpha \beta}(t)= \langle [c_{\alpha}(t)c_{\beta}(t), H(T,U)]\rangle.
\end{equation}
As $H(T,U)$ involves terms quartic in the Majorana operators this
evolution clearly takes us out of the Gaussian setting. We
truncate this transformation imposing that Wick's theorem holds,
i.e. we write
\begin{equation}
\langle c_i c_j c_k c_l \rangle = -(\Gamma_{ij}\Gamma_{kl} -
\Gamma_{ik}\Gamma_{jl} + \Gamma_{il}\Gamma_{jk}).
\end{equation}
With the help of the commutation relation
\begin{eqnarray*}
\left[c_{\alpha}c_{\beta}, c_i c_j\right] &=& 2(c_i
c_{\alpha}\delta_{j\beta} - c_i c_{\beta}\delta_{j \alpha} -
c_{\beta} c_{j} \delta_{i \alpha} + c_{\alpha}c_j \delta_{i
\beta}),
\end{eqnarray*}
we can calculate the contributions for the quadratic term $H_q = \sum_{kl}T_{kl}c_kc_l$ and the
interaction term $H_I = \sum_{klmn}U_{klmn}c_kc_lc_mc_n$ to be
\begin{eqnarray*}
\langle [c_{\alpha} c_{\beta}, H_q] \rangle &=& 4
[T,\Gamma]_{\alpha
\beta},\\
\langle [c_{\alpha} c_{\beta}, H_I] \rangle &=&
24[\tr_B[U \Gamma],\Gamma]_{\alpha\beta}.
\end{eqnarray*}
This implies the following time evolution of the covariance matrix:
\begin{eqnarray}
\frac{d}{dt}\Gamma(t) &=& 4
[\bar{h}(\Gamma(t)), \Gamma(t)],\label{eq:Gammadot}\\
\bar{h}(\Gamma(t)) &=& T+6\tr_B[U\Gamma(t)],\label{eq:hbar}
\end{eqnarray}
where we have defined $\tr_B[U\Gamma]_{ij} \equiv \sum_{kl}U_{ijkl}\Gamma_{lk}$. This equation can be
formally integrated:
\begin{eqnarray}
\Gamma(t) &=& O(t)\Gamma(0)O(t)^T, \nonumber\\
O(t) &=& \mathcal{T}\exp \left( 4\int_0^t \bar{h}(\Gamma(t')) dt' \right), \label{eq:Ot}
\end{eqnarray}
where $\mathcal{T}$ denotes the time ordering operator. Note that due to the anti-symmetry of
$\bar{h}(\Gamma(t'))$ Eq. \eqref{eq:Ot} guarantees that the matrix $O(t)$ is an orthogonal
transformation. Hence, when starting with a valid covariance matrix $\Gamma(0)$, this approximation
scheme ensures that we remain within the set of Gaussian states. Further, it follows from what we have said in
Sec.~\ref{sec:methods} that the dynamical evolution under the interaction Hamiltonian
$H(T,U)$ can be understood as the time evolution under a \emph{quadratic but state-dependent} Hamiltonian
\begin{eqnarray}
H_Q(\Gamma) &=& i\sum_{kl} \bar{h}(\Gamma(t))_{kl}c_kc_l.
\end{eqnarray}
Further, this approximation scheme does not only ensure that we always remain within the set of Gaussian states,
but a short calculation shows that it also conserves energy and, for a number conserving Hamiltonian, the particle number. We start with the energy, which is given by $E(t) = \tr[H\rho] = -\tr[(T + 3 \tr_B[U\Gamma])\Gamma]$, and hence, with the help of Eq.~\eqref{eq:Gammadot}, it follows that 
\begin{align}
\dot E (t) = \sum_{kl}\frac{\partial E}{\partial \Gamma_{kl}}\dot \Gamma_{kl} = 4 \tr[÷\bar h (\Gamma) [\bar h (\Gamma), \Gamma]] = 0. 
\end{align}
To prove the conservation of the mean particle number $N$, as long as $[H, \hat N] = 0$, we write the particle number operator $\hat N = \sum_k \ad_k a_k = \tfrac{M}{2} + \tfrac{i}{4} \sum_{k,l}\nu_{kl} c_kc_l$, where $\nu$ is real and anti-symmetric, and obtain $N(t) = \tfrac{M}{2} - \tfrac{1}{4}\tr[\nu \Gamma]$. Thus (c.f. Eq. \eqref{eq:Gamma_dot_commutator})
\begin{align*}
\frac{d}{dt} N(t) = - \frac{1}{4}\tr[\nu \dot\Gamma] =  - \frac{1}{4}\sum_{kl} \nu_{kl} \tr[\rho [c_lc_k, H]]\\
=  -i \tr[\rho [\hat N, H]] =0.
\end{align*}

Hence, we have shown
that applying Wick's theorem to the evolution equation of a system governed by a two-body
interaction Hamiltonian leads to a consistent dynamical equation for the CM. The truncation of the evolution
is equivalent to an evolution under an effective state-dependent quadratic Hamiltonian within the set of Gaussian states, as depicted in
Fig. \ref{fig:Scheme} a). We remark that this approach works for pure as well as mixed FGS. Note also that the remaining equation resembles in some way the
Gross--Pitaevskii equation for Bosons since that one can also be interpreted as giving the evolution in terms of a Hamiltonian that depends on the state; however, the main difference is that in the case of Bosons such an equation is obtained for the first moments of the field operators. Note that in the case of Bosons one could also perform the same 
approximation for the first and second moments, obtaining a set of coupled equations which would include the condensate part as well as thermal particles.

\subsection{Ground States}\label{sec:ground_state}
In principle, the ground state in gHFT can be found via a direct minimization
\begin{multline}\label{eq:min-E-Gauss}
\min_{\rho \;\mathrm{Gaussian}} \tr[H\rho] = \\ \min_{i\Gamma \leq \mathds{1} }\sum_{ij}
\left\{T_{ij}\Gamma_{ij} - 3\sum_{i,j,k,l}U_{ijkl}\Gamma_{ij}\Gamma_{kl}\right\}.
\end{multline}
For the translationally invariant Hubbard model this problem could be reduced to an optimization over
two parameters only \cite{bach-1993}. Generically, this constrained quadratic minimization problem is a
daunting task. We attack this problem from a different perspective. A well-known approach for the
determination of the ground state $|\phi_0\rangle$ of a Hamiltonian $H$ is imaginary time evolution.
Due to the exponential growth of the state space with the number of modes this approach can be applied
to small systems only. The idea is to apply the Gaussian approximation (in the form of Wick's theorem) to derive
an evolution equation for the CM. In this way we get a simple quadratic scaling in the computational time with respect to the number of
modes. 

Naively, one would think of obtaining the evolution equation in imaginary time by replacing $t \mapsto it $ in the real-time evolution equation of the CM, Eq. ~\eqref{eq:Gammadot}. However, as it turns out, this is not the right road to take.  Though, as we will show below, the following approach will lead us to the desired ground state: We start with an arbitrary pure Gaussian state $\rho(0)$, and evolve it according to
\begin{equation}\label{eq:rho_t_imag}
\rho(t) = \frac{e^{-Ht}\rho(0)e^{-Ht}}{\tr[e^{-2Ht}\rho(0)]},
\end{equation}
where the normalization of the density operator for any time, i.e. the denominator $\tr[e^{-2Ht}\rho(0)]$, is crucial for obtaining the correct equations. Since we apply the non-quadratic operator $H$ to the Gaussian state $\rho$, this procedure clearly takes us out of the setting of Gaussian states. 
One way around would consists of discretizing the evolution and finding
for small time steps $\Delta t$ the best Gaussian approximation at each step. However, due to the
truncation it is not clear that this procedure will converge. And even if we find a steady state it is
not clear that this will be the ground state of the system. Another possible approach is to use the
quadratic but state-dependent effective Hamiltonian $H_Q(\Gamma)$ that is derived from $H(T,U)$ for the
imaginary time evolution. As $H_Q(\Gamma)$ is a quadratic operator we will stay in the space of
Gaussian states. But as the Hamiltonian is state-dependent the outcome of this procedure is not
clear. However, as we will show below, both approaches are equivalent and lead indeed to the
desired solution. To be precise, we show that the following is equivalent:
\begin{enumerate}
 \item Direct minimization of the energy Eq.~\eqref{eq:min-E-Gauss} in generalized Hartree-Fock theory.
 \item Imaginary time evolution of the state $\rho$ with the full Hamiltonian $H(T,U)$ for small time steps $\Delta t$ followed by
 an approximation of $\rho(t + \Delta t)$ by a Gaussian state.
 \item Imaginary time evolution of $\rho$ with the quadratic but state-dependent Hamiltonian
 $H_Q(\Gamma)$.
\end{enumerate}
%
\subsubsection{Minimization of the energy} In order to obtain the generalized Hartree-Fock ground state we have
to solve the minimization problem
\begin{multline}\label{eq:minE}
\min_{\rho \;\mathrm{Gaussian}}E(\rho) = \min_{\rho \;\mathrm{Gaussian}} \tr[H\rho] = \\\min_{i\Gamma
\leq \mathds{1} }\left\{\sum_{ij} T_{ij}\Gamma_{ij} -
3\sum_{i,j,k,l}U_{ijkl}\Gamma_{ij}\Gamma_{kl}\right\}.
\end{multline}
It has been proven in Ref.~\cite{bach-1993} that the HF ground state is always pure, i.e. $\Gamma^2 =
-\mathds{1}$. Using Lagrange multipliers, we arrive at the following necessary conditions for a local
minimum (see Appendix \ref{app:Minimize_energy_Lagrange})\footnote{Note that Eqs.~\eqref{eq:hbar-Gamma} and ~\eqref{eq:G_pure} describe all minima, local and global ones. However, for the particular problems studied in the next sections, our numerical investigations show that our algorithms lead in general to the global minimum, i.e. the ground state.}:
\begin{eqnarray}
[\bar{h}(\Gamma),\Gamma] &=&0, \label{eq:hbar-Gamma}\\
\Gamma ^2 &=& - \mathds{1} \label{eq:G_pure}.
\end{eqnarray}
These two equations are non-linear matrix equations and thus hard to solve, both analytically and
numerically, for large systems. But we show next that these equations appear as the steady-state
conditions of imaginary time evolution.
%
\subsubsection{Imaginary time evolution}
From Eq. \eqref{eq:rho_t_imag} we see that the evolution of the density operator $\rho$ under any
Hamiltonian $H$ in imaginary time is given by
\begin{equation}
\dot{\rho}(t) = -\{H,\rho(t)\}+ 2\rho(t)\tr[H\rho(t)],
\end{equation}
so that the covariance matrix evolves according to
\begin{equation}\label{eq:Gammadot_imag_equ}
\frac{d}{dt}\Gamma_{kl}(t) = - i\tr[\{H,c_kc_l\}\rho(t)] + 2\Gamma_{kl}\tr[H\rho(t)].
\end{equation}
We show in Appendix \ref{app:Proof_imaginary_time} that both approaches for imaginary time evolution, number 2 and 3,
lead to the same evolution equation of the covariance matrix:
\begin{equation}\label{eq:Gammadot_imag}
\frac{d}{dt}\Gamma = -4 \left(\Gamma \bar{h}(\Gamma) \Gamma + \bar{h}(\Gamma)\right).
\end{equation}
First, note that Eq. \eqref{eq:Gammadot_imag} ensures that a pure state remains pure under the evolution, since
\begin{eqnarray*}
\frac{d}{dt}\Gamma^2 = \Gamma \dot \Gamma + \dot \Gamma \Gamma =-4(\Gamma^2\bar h \Gamma + \Gamma \bar h + \Gamma \bar h \Gamma^2 + \bar h \Gamma) = 0, 
\end{eqnarray*}
as all pure states fulfill $\Gamma^2 = -\mathds{1}$. Next, we show that the energy always decreases under the evolution:
\begin{multline}
\frac{d}{dt}E(t) =\sum_{kl} \frac{\partial E}{\partial \Gamma_{kl}}\dot \Gamma_{kl} = \sum_{kl}\bar h(\Gamma)_{kl}\dot \Gamma_{kl}\\
= 4\tr[\bar h (\Gamma \bar h \Gamma +\bar h)]  = 2 \tr[([\bar h, \Gamma])^2] \leq 0. \label{eq:E_decreases}
\end{multline}
Here, we have used that the matrix $[\bar h, \Gamma]$ is antisymmetric, implying that $([\bar h, \Gamma])^2$ is negative-definite. Further, it follows from Eq. \eqref{eq:E_decreases} that the steady-state is reached iff $[\bar h, \Gamma] = 0$. Thus, we are ensured to reach the gHF ground state via an evolution with Eq.  ~ \eqref{eq:Gammadot_imag} with initial condition $\Gamma(0)^2 = -\mathds{1}$. 

Finally, if we start in a pure input state, i.e. $\Gamma(0)^2 = -\mathds{1}$, Eq.  ~ \eqref{eq:Gammadot_imag} can be formally integrated (see also Fig.~\ref{fig:Scheme}b ):   
\begin{eqnarray}
\Gamma(t) &=& O(t)\Gamma(0)O(t)^T, \nonumber\\
O(t) &=& \mathcal{T} \exp \left[\int_0^t A(\Gamma(t'))dt'\right],\label{eq:O_imag}\\
A(\Gamma(t)) &=& 2[\bar{h}(\Gamma), \Gamma].
\end{eqnarray}
The evolution of $\Gamma$ with the orthogonal $O(t)$ has turned out to be especially fast and stable in numerical implementations compared to the direct integration of Eq.  ~ \eqref{eq:Gammadot_imag} via e.g. a Runge--Kutta solver, and we have used it for our numerical applications.

\subsection{Thermal states}\label{sec:thermal_state}
As we have seen in the last Section, the generalized Hartree-Fock ground state can be obtained via an
imaginary time evolution. To learn about the finite temperature properties in gHFT we have to consider
the approximation to the Gibbs state $\rho \sim e^{-\beta H}$, where $\beta = \tfrac{1}{k_BT}$ is the
inverse temperature. We recall that the Gibbs state minimizes the free energy, $F(\rho) = E(\rho) -
\beta^{-1} S$. Thus, we have to solve
$$
\min_{\rho \;\mathrm{Gaussian}} F(\rho) = \min_{\rho \;\mathrm{Gaussian}} \left\{E(\rho) -
\beta^{-1}S(\rho)\right\}.
$$
As in the case of the minimization of the energy, we reformulate the problem as an optimization problem using Lagrange multipliers.  As we show in Appendix \ref{app:Minimize_Free_Energy}, this approach leads to the following necessary conditions for a minimum of the free energy:
\begin{eqnarray}
[h_F(\Gamma),\Gamma] &=& 0, \label{eq:HFcom}\\
h_F(\Gamma) (\mathds{1} + \Gamma^2) &=& 0, \label{eq:HF0}
\end{eqnarray}
where
\begin{equation}
h_F(\Gamma) = \bar{h} - \frac{i}{4\beta}\ln \frac{\mathds{1} + i \Gamma}{\mathds{1} - i \Gamma}.
\end{equation}
Now, assume for the moment that $\det(\mathds{1}+\Gamma^2) \neq 0$. Then, according to Eq. \eqref{eq:HF0}, $h_F = 0$ is the only possible solution. Hence, the CM has to be of the form
\begin{equation}\label{eq:Gamma_Gibbs}
\Gamma = i \tanh \left[2i\beta\bar{h}(\Gamma)\right].
\end{equation}
Now, a CM of this form always fulfills Eqs. \eqref{eq:HFcom} and \eqref{eq:HF0}, and includes, in the limit $\beta \rightarrow \infty$, the zero-temperature case, so that Eq.~\eqref{eq:Gamma_Gibbs} is indeed a solution of Eqs. \eqref{eq:HFcom} and \eqref{eq:HF0}.

Note that from the perspective of numerical implementation, we can solve Eq. \eqref{eq:Gamma_Gibbs} e.g. via a fixed-point iteration (see Fig.~\ref{fig:Scheme}c). 

\begin{figure}[t]
\begin{center}
\includegraphics[width=0.95\columnwidth]{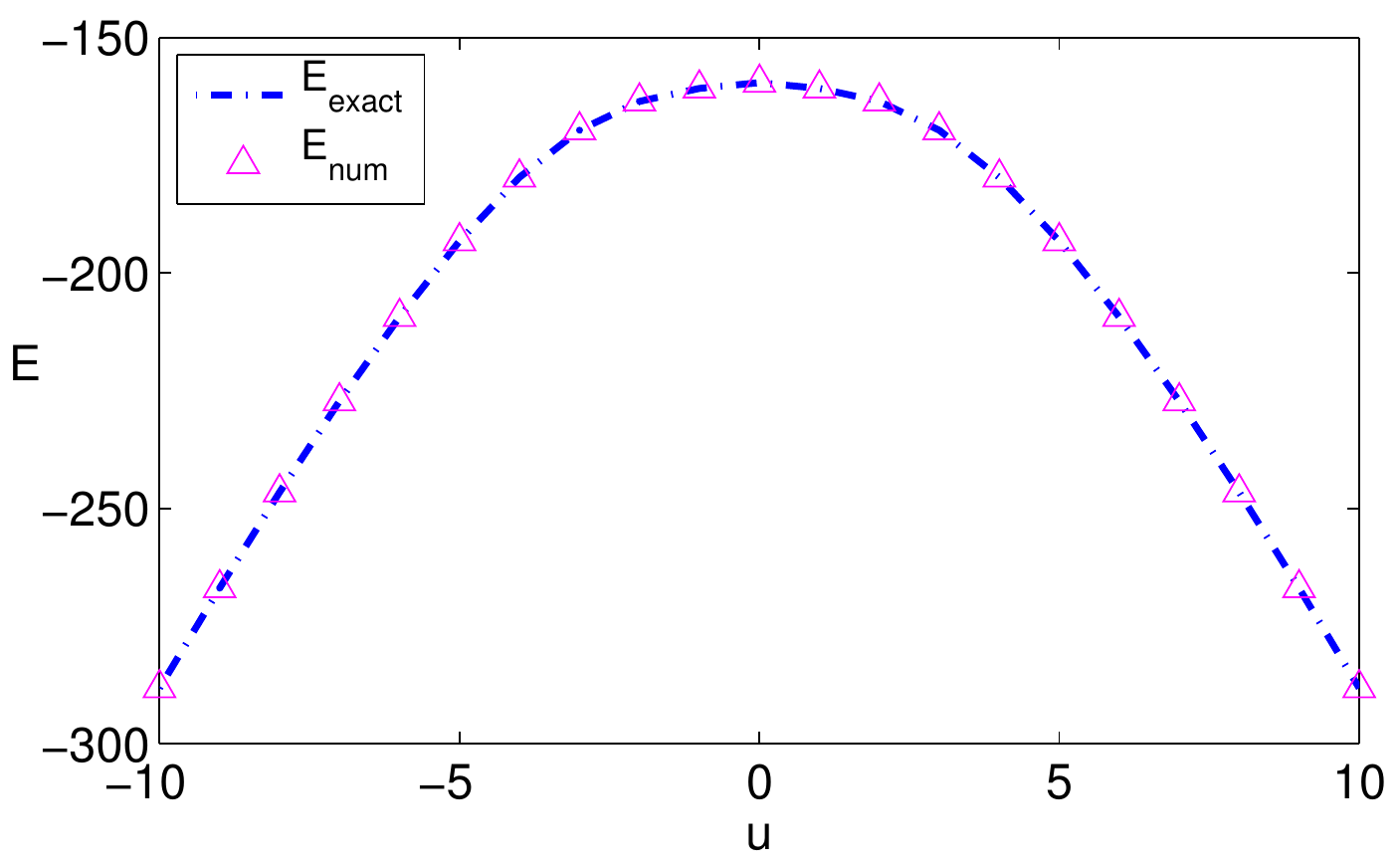}
\end{center}
\caption{ \label{Lieb_vs_exact_Nh10} Ground state energy of the translationally invariant Hubbard
model (c.f. Eq. \eqref{eq:Hubbard_H_2d}) for an interaction strength $u \in [-10,10]$ at half-filling, i.e. $\mu = 0$.
The blue line shows the exact gHF ground state energy, $E_{\mathrm{exact}}$, obtained from
Ref.~\cite{bach-1993}, while the triangles correspond to the energy obtained via imaginary time
evolution, $E_{\mathrm{imag}}$. }
\end{figure}

\section{Benchmark: The two-dimensional translationally invariant  Hubbard model}\label{sec:numerics}
In the following Section we apply our method to the two-dimensional translationally invariant Hubbard model with periodic boundary conditions:
\begin{multline}\label{eq:Hubbard_H_2d}
H= t\sum_{ \langle x,y  \rangle \in \Lambda, \sigma}\ad_{x,\sigma}a_{y,\sigma} \\+ u \sum_{x\in
\Lambda} \left(n_{x\uparrow}-\frac{1}{2}\right)\left(n_{x\downarrow}-\frac{1}{2}\right) + \mu
\sum_{x,\sigma}n_{x,\sigma},
\end{multline}
where $x$ and $y$ are points on a two-dimensional lattice, $\langle i,j \rangle$ denote
nearest-neighbors and $\sigma = \uparrow, \downarrow$ denotes the spin degree of freedom. We consider only the case where we have the same number of spin-up and spin-down particles, so that we can use the same chemical potential for the two species. For $u < 0$
$(u > 0 )$ the second term in $H$ is an attractive (repulsive) on-site interaction between particles of
opposite spin. In the following, we set $t=1$ as the energy scale of the system. Note that the case of half-filling is characterized by $\mu = 0$.

The goal of this Section is two-fold: In the first part, Sec. \ref{sec:benchmark}, we benchmark our numerical method by considering physical quantities for which an exact solution within gHFT is known. In the case of the translationally invariant Hubbard model it could be proven in Ref. ~\cite{bach-1993} that the (free) energy for the (thermal) ground state can be found via a two-parameter optimization, and we will compare our numerical results with the numbers obtained from the optimization. After the demonstration of the power of our approach, we continue in the second part, Sec.~\ref{sec:Compare_QMC}, with a comparison of gHFT itself to more powerful and and sophisticated methods, like Quantum Monte Carlo (QMC). 

In the following, all calculations are performed for a $10 \times 10$ lattice.

\begin{figure}[t]
\begin{center}
\includegraphics[width=0.95\columnwidth]{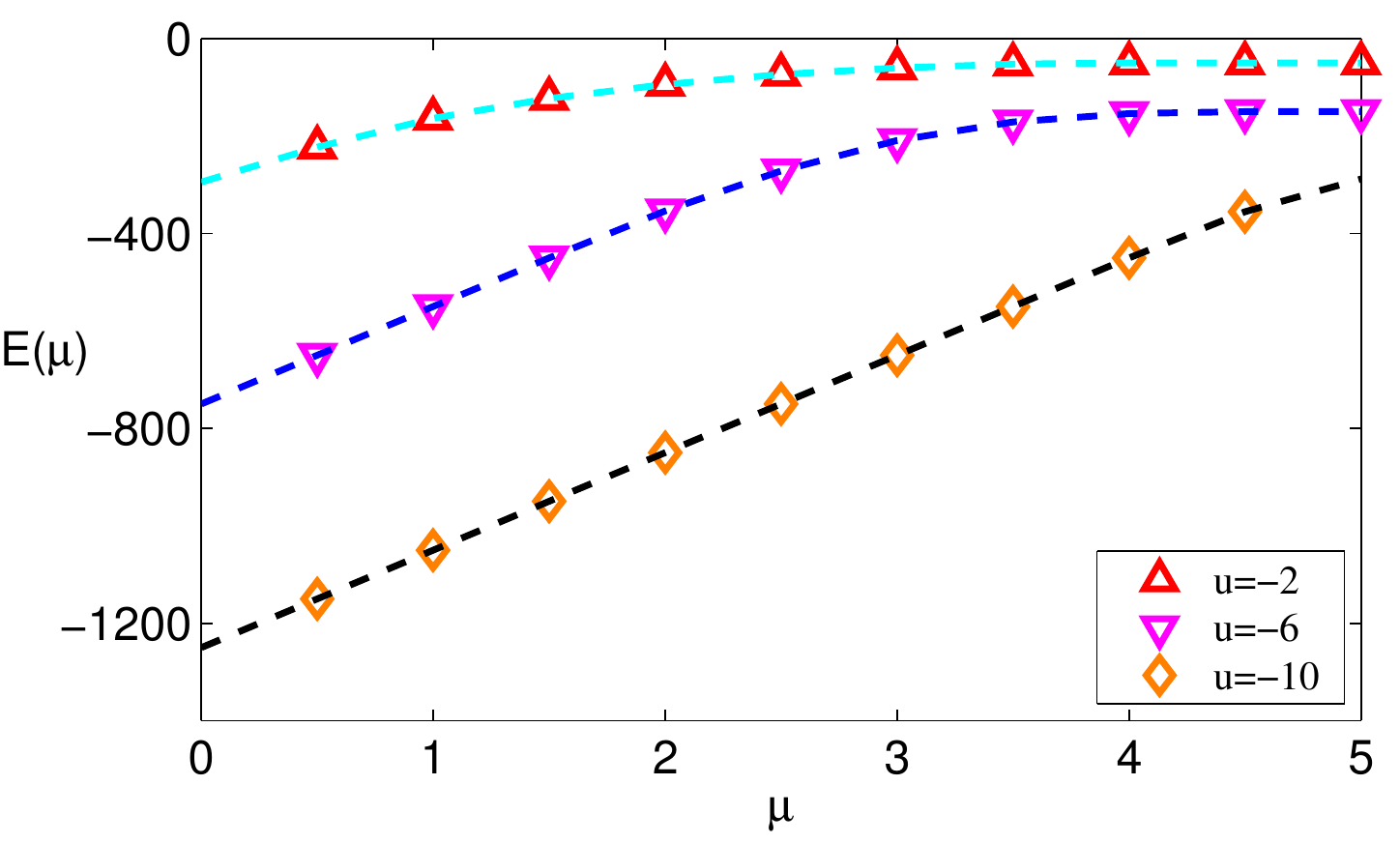}
\end{center}
\caption{Energy as a function of the chemical potential for the attractive Hubbard model at various values of the interaction strength $u$. The dotted lines depict the exact gFH solution, while the triangles and diamonds correspond to the solutions obtained via our approach. We find excellent agreement of the two approaches \label{fig:Energy_mu}}
\end{figure}

\subsection{Comparison with the exact gHF solution}\label{sec:benchmark}
For the translationally invariant Hubbard model Eq.~\eqref{eq:Hubbard_H_2d} the exact results for the energy of ground and thermal state within gHFT where presented in ~\cite{bach-1993}. In Fig.~\ref{Lieb_vs_exact_Nh10} we compare the ground state energy for half-filling obtained via
imaginary time evolution according to Eq.~\eqref{eq:O_imag} (triangles)
with the exact gHF solution (blue line) and find excellent agreement. The relative error
in the energy is given by $(dE)_{\mathrm{rel}}= |E_{\mathrm{imag}} -
E_{\mathrm{exact}}|/|E_{\mathrm{exact}}|< 2\cdot 10 ^{-8}$.

In Fig.~\ref{fig:Energy_mu} we show results away from half-filling. In this case, a closed solution for the energy could only be provided for the attractive Hubbard model in Ref.~\cite{bach-1993}. We have compared the exact gHF solution (dotted lines) with the results obtained via our approach (triangles and diamonds) for $u=-2, -6, -10$, proving that our numerical method also works well for a doped system.

\begin{figure}[t]
\begin{center}
\includegraphics[width=0.95\columnwidth]{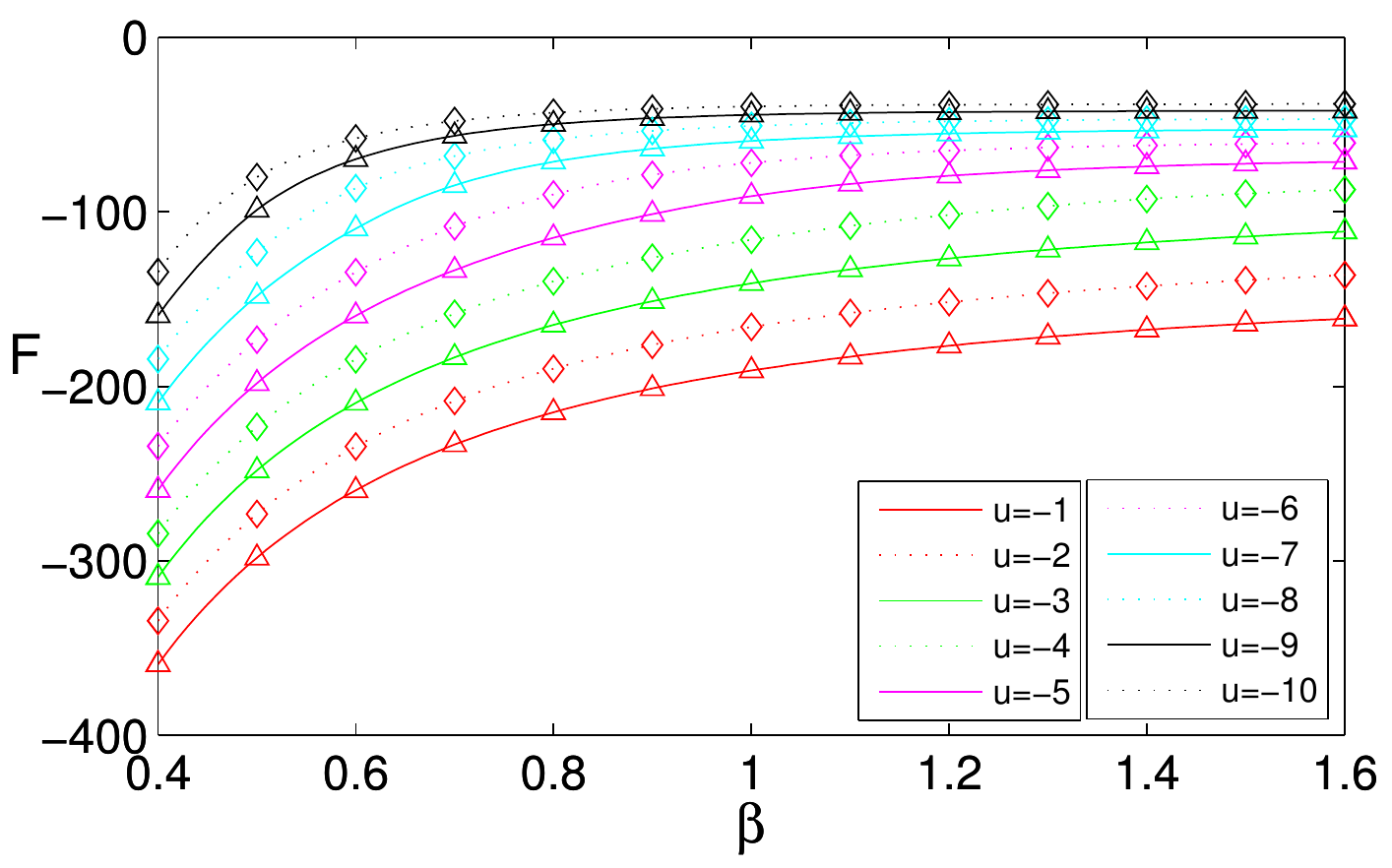}
\end{center}
\caption{Free energy of the translationally invariant Hubbard model (c.f. Eq.
\eqref{eq:Hubbard_H_2d}) in the temperature range $\beta \in [0.4,1.6]$
for $u = -1, \ldots, -10$ at half-filling. The lines depict the gHFT solution given in Ref.~\cite{bach-1993}, while the
triangles and diamonds show the results obtained via the fixed-point iteration.\label{fig:free_energy}}
\end{figure}

Next, we consider the case of finite temperature, where the gHF Gibbs state is given by the implicit
equation~\eqref{eq:Gamma_Gibbs}. Starting from
the ground state we change the temperature in steps $\Delta \beta = 0.01$ and compute the new
thermal state via a fixed-point iteration. In Fig.~\ref{fig:free_energy} we present a comparison of our
solutions (triangles and diamonds) for half-fiiling with the exact results (lines) for various
values of $u$ in a temperature range of $\beta \in [0.4, 1.6]$, and find excellent agreement.

\begin{figure}[t]
\begin{center}
\includegraphics[width=0.95\columnwidth]{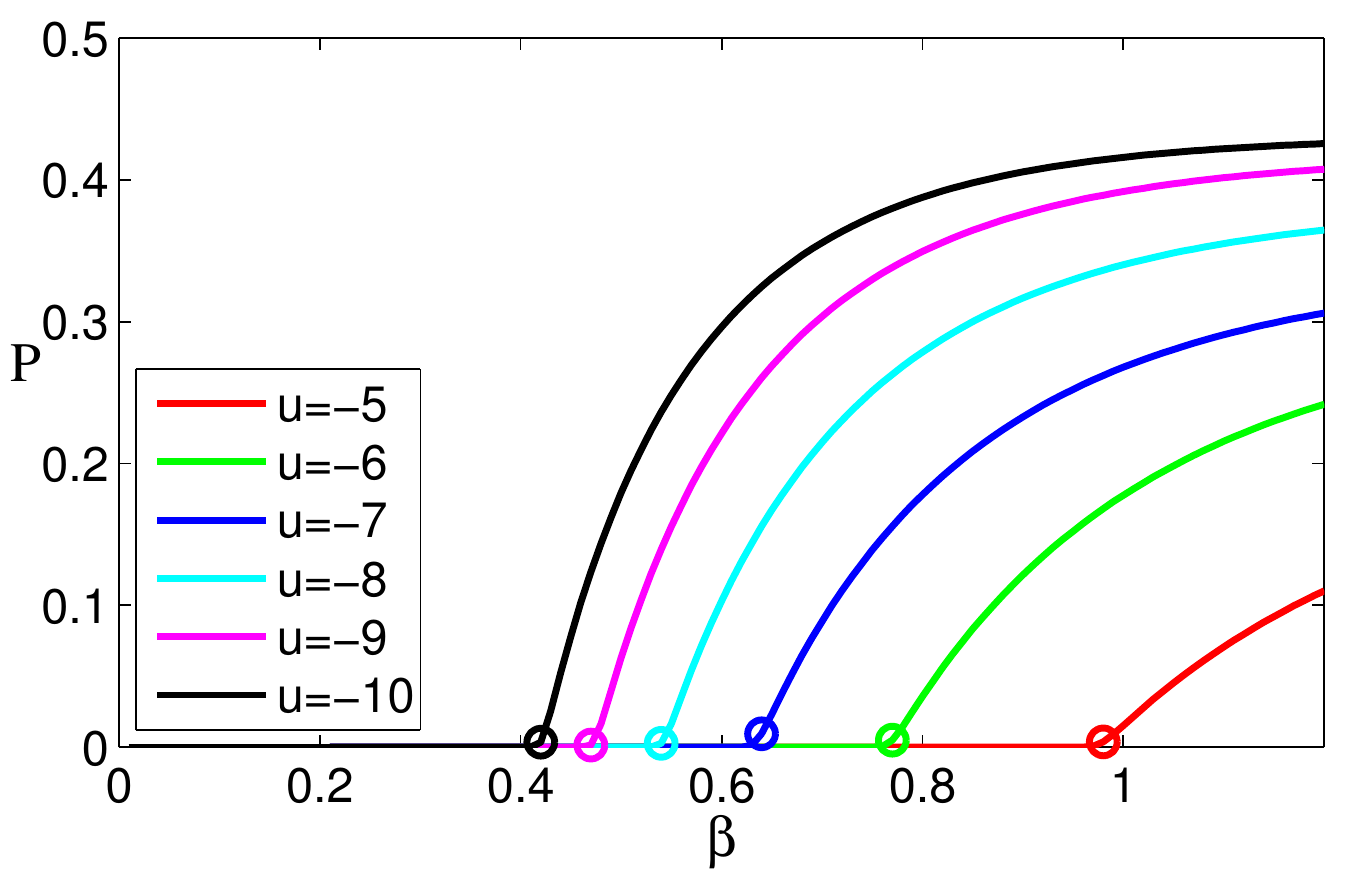}
\end{center}
\caption{Temperature-dependence of the pairing (see  Eq.~\eqref{eq:Def_Pairing}) for various values of the interaction $u$ at half-filling. We observe
the occurrence of a phase transition from a paired phase to an unpaired normal state.
The circles indicate the value of the transition temperature derived for gHFT in
Ref.~\cite{bach-1993}.\label{fig:phase_transition}}
\end{figure}

One of the most interesting questions concerning the Hubbard model is the appearance of 
superconductivity. As we have already pointed out in Sec.~\ref{sec:statement}, FGS describe among others superfluid and normal states, since every pure state of this family can be
brought into the standard (BCS) form $|\Psi\rangle = \prod_k (u_k + v_k \ad_k\ad_{-k})\vac$ via a
change of basis. In this basis, the gap parameter $\Delta = \sum_k u_k v_k$ is an order parameter for
the superfluid phase. To consider superfluidity independent of the basis, it is more appropriate to
consider the normalized pairing measure derived in \cite{kraus:022303}:
\begin{equation}\label{eq:Def_Pairing}
P = \frac{2}{M}\sum_{kl}|\langle \ad_k \ad_l\rangle|^2.
\end{equation}
Then, a positive value of $P$ indicates that we are in a paired phase, while a normal state is
characterized by $P=0$. 

It has been shown in Ref.~\cite{bach-1993} that within gHFT the attractive Hubbard model exhibits a phase 
transition from a paired phase to an unpaired normal phase at a critical temperature $\beta_c$  for any finite lattice size, and the corresponding transition temperature has been derived. In Fig.~\ref{fig:phase_transition} we have depicted the pairing as a function of the inverse temperature for various values of the interaction $u$,  and find a breakdown at a finite value $\beta_c$ that depends on the interaction strength. The values of $\beta_c$ that we obtain via our numerical method agree well with the results presented in Ref.~\cite{bach-1993}, and that are depicted as circles in  Fig.~\ref{fig:phase_transition}.  Further, we give a list of the critical exponent $\gamma$ defined via $P = a (T_c -T)^{\gamma}$ for various values of $u$ in table~\ref{Crit_P}.

\begin{table}[htdp]
\begin{center}
\begin{tabular}{c|c|c|c|c|c|c|}
\hspace{0.1cm}u \hspace{0.1cm} &$-5$ & $-6$ & $-7$ & $-8$ & $-9$ & $-10$\\\hline
\hspace{0.1cm} $ \gamma$ \hspace{0.1cm} & \hspace{0.1cm} 0.87 \hspace{0.1cm}&  \hspace{0.1cm} 0.90  \hspace{0.1cm} 
 &  \hspace{0.1cm} 0.80  \hspace{0.1cm} &\hspace{0.1cm} 1.04 \hspace{0.1cm} & 
 \hspace{0.1cm}1.14 \hspace{0.1cm}& \hspace{0.1cm}1.00\hspace{0.1cm}
\end{tabular}
\end{center}
\caption{Critical exponent $\gamma$ for the phase transition from a paired to an unpaired phase for various values of u. We have fitted $P$ around the critical temperature $T_c$ to a curve of the form $P = a (T_c -T)^{\gamma}$ for lattice sizes $8\times8$, $10\times10$ and $12\times12$, finding the same values of $\gamma$ for the given precision. \label{Crit_P}}
\end{table}%

\begin{figure}[t]
\begin{center}
\includegraphics[width=0.95\columnwidth]{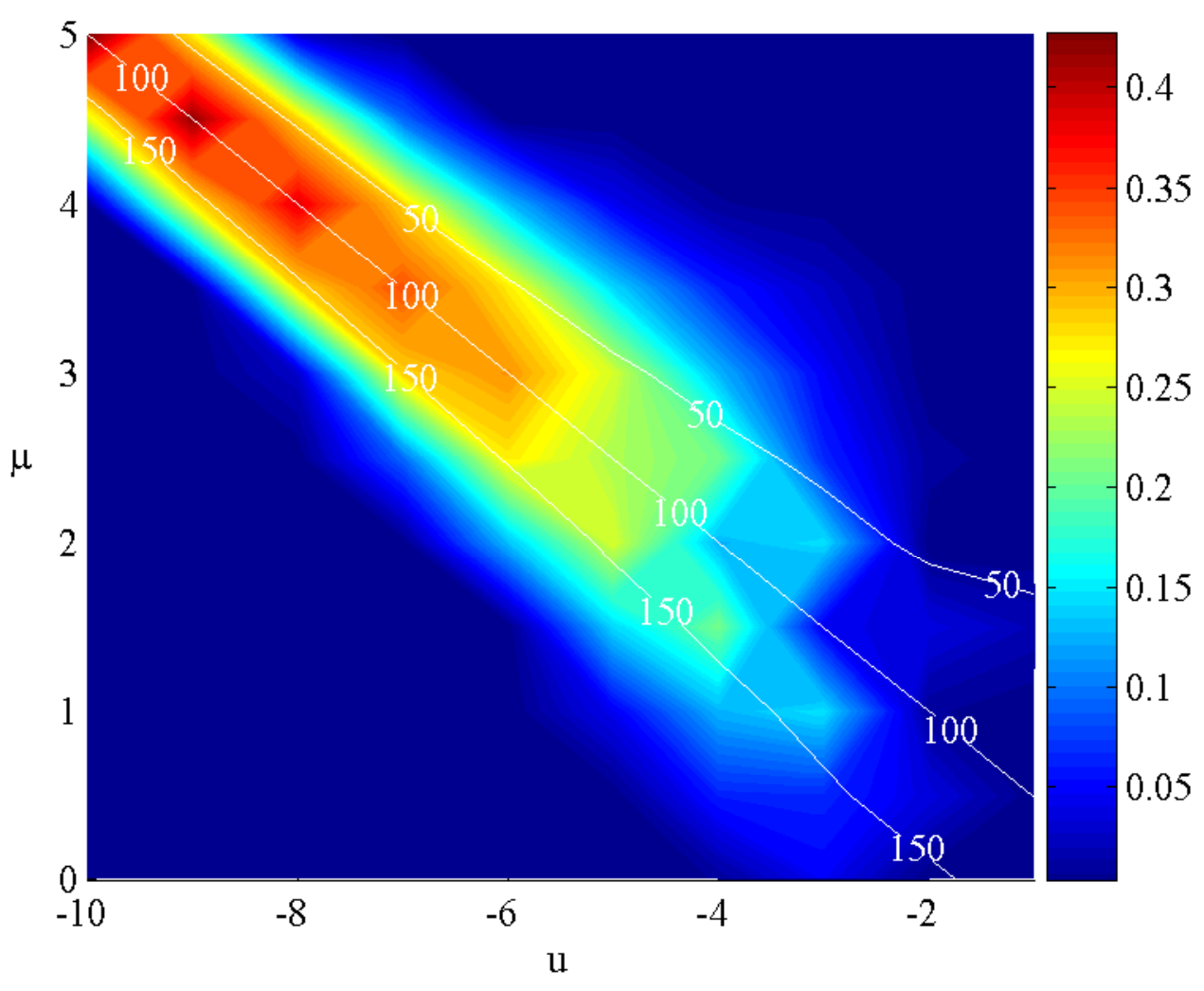}
\end{center}
\caption{Pairing (see Eq.~\eqref{eq:Def_Pairing}) of the ground state as a function of the interaction strength and the chemical
potential in the translationally invariant case. As a guide for the eye, we have added lines for the particle
numbers $50$, $100$ (half filling) and $150$. \label{fig:Phase_diagram_T0}}
\end{figure}

Finally, we give in Fig. \ref{fig:Phase_diagram_T0} a phase diagram of the pairing for the ground state as a function of the chemical potential and the interaction strength. (Note that the value of the pairing is not unique in the case of half-filling due to the Gauge symmetries of the Hamiltonian (see Ref.~\cite{bach-1993}). We have restricted to represent the phase diagram for negative values of $u$, since we find that $P = 0$ for $u$ positive.

In summary, we have demonstrated that our numerical methods are capable of obtaining the correct gHF ground and Gibbs state for the translationally invariant Hubbard model, both in the attractive as well as in the repulsive regime. Having demonstrated the validity of our approach, the next obvious question is now how gHFT itself compares to more powerful and sophisticated methods, like Quantum Monte Carlo (QMC).

\subsection{Comparison with Quantum Monte Carlo}\label{sec:Compare_QMC}
In this Section we compare the results obtained via our algorithm to numerical data from QMC simulations.  To this end, we will concentrate on the repulsive Hubbard model at half-filling. The reason for this choice is two-fold: First, as it has been shown in  \cite{Micnas_RMP, Scalapino_BCS}, the physics of the attractive Hubbard model is well-described by BCS-theory, which is included in gHFT. This motivates to consider primarily the repulsive Hubbard model, and compare our results with results obtained from QMC. While QMC suffers from the notorious sign problem when applied to fermionic systems, it could be shown that in the case of a half-filled lattice the sign problem does not occur \cite{Muramatsu_sign}, and thus QMC becomes exact. As a consequence, the two-dimensional repulsive Hubbard model at half-filling has undergone an intense investigation in recent years. A partial list of these results is given in ~\cite{Giamarchi, Hirsch, White_QMC_94, White_QMC_95, Zhang_QMC, Dagotto_QMC, Rozenberg_QMC, Evertz_QMC, Hanke_QMC, Hettler_QMC, Tremblay_QMC} and references therein. In this Section, we compare our results obtained in gHFT with the recent QMC results obtained in Ref.~\cite{Scarletter} and Ref.~\cite{trivedi-2009} for half-filled and doped lattices. 

\subsubsection{Ground state properties for half-filling}
We start with a comparison of our approach with the results presented in Ref.~\cite{Scarletter}, where a half-filled system is considered. Since in the latter work the numerical results are presented for various lattice sizes, while our results are always given for a $10 \times 10$ lattice, we have to restrict to a qualitative comparison in certain cases. As the first physical quantity, we consider the momentum distribution $n(k)$ that has been depicted for $u = 2, 3, 4, 5, 6, 8$ and very small, but non-zero temperature in  Ref.~\cite{Scarletter}. In Fig. \ref{fig:Momentum_Distribution} we show our results for the momentum distribution $n(k)$ for an interaction $u = 1, \ldots, 10$ at half-filling in the entire Brioullin zone. Our results show an excellent qualitative agreement with  Fig.~1a of Ref.~\cite{Scarletter}. As in Ref ~\cite{Scarletter} we find that with increasing $u$, the momentum distribution fulfills $n(0,0) \rightarrow 1$ and $n(\pi,0) = 1/2$ for all values of the interaction strength, and $n(k)$ grows with increasing $u$ for $ k \in [(\pi, \pi), (\pi,0)]$.  Further, as in ~\cite{Scarletter} we observe a broadening of the Fermi surface with growing interaction strength, i.e. the momentum distribution gets smeared out.

\begin{figure}
\begin{center}
\includegraphics[width = 0.9\columnwidth]{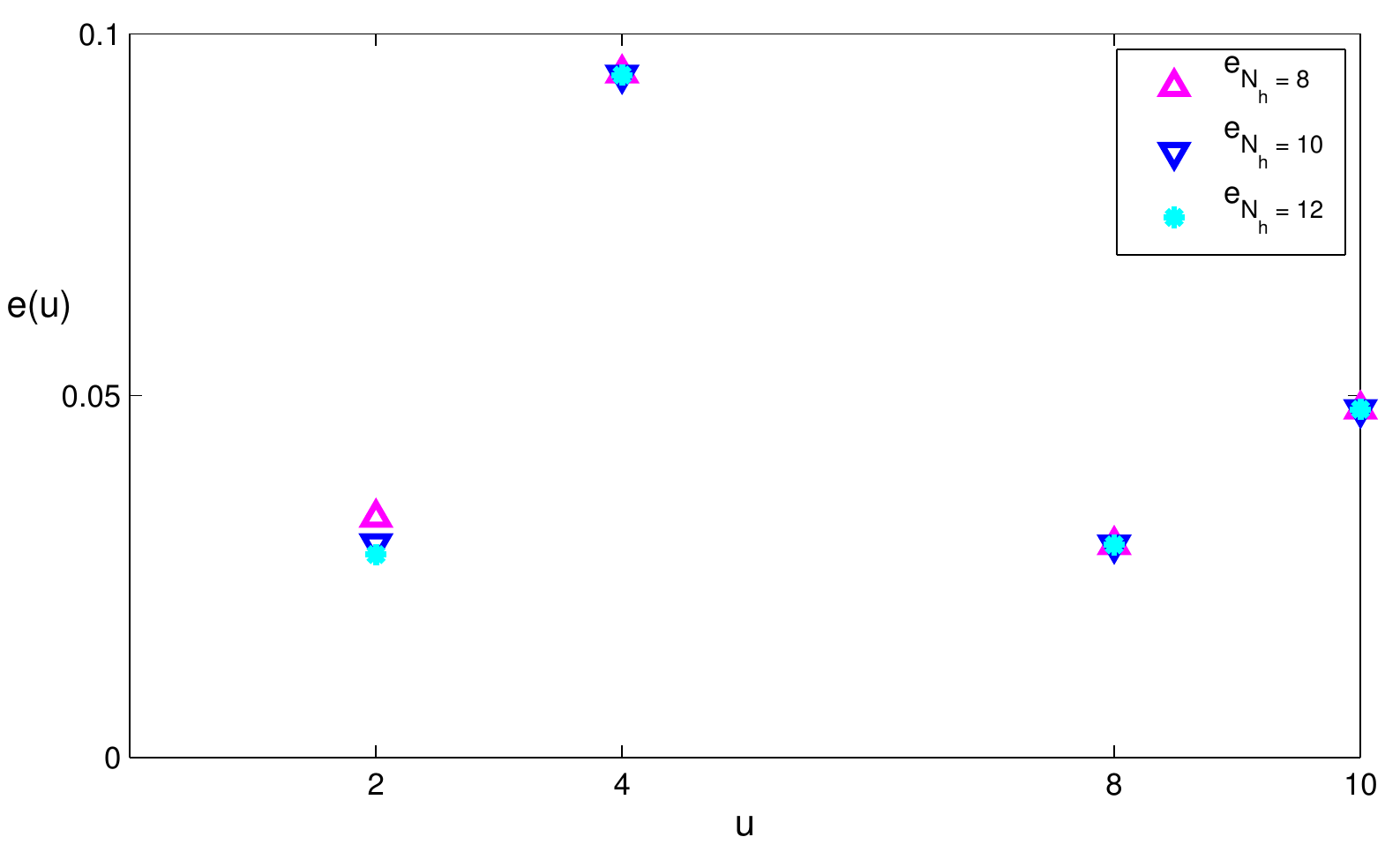}
\end{center}
\caption{Comparison of the ground state energy obtained from gHFT with the QMC results of Refs.~\cite{Giamarchi} and ~\cite{Hirsch}. We have depicted the relative error $e(u) = |E_{QMC}(u) - E_{gHFT}(u)|/|E_{QMC}(u)| $ for lattice sizes $8 \times 8$, $10 \times 10$ and $12 \times 12$ and interaction strengths $u = 2, 4, 8, 10$.
\label{fig:Compare_GS_energy_QMC} }
\end{figure}

\begin{figure}
\begin{center}
\includegraphics[width = 0.95\columnwidth]{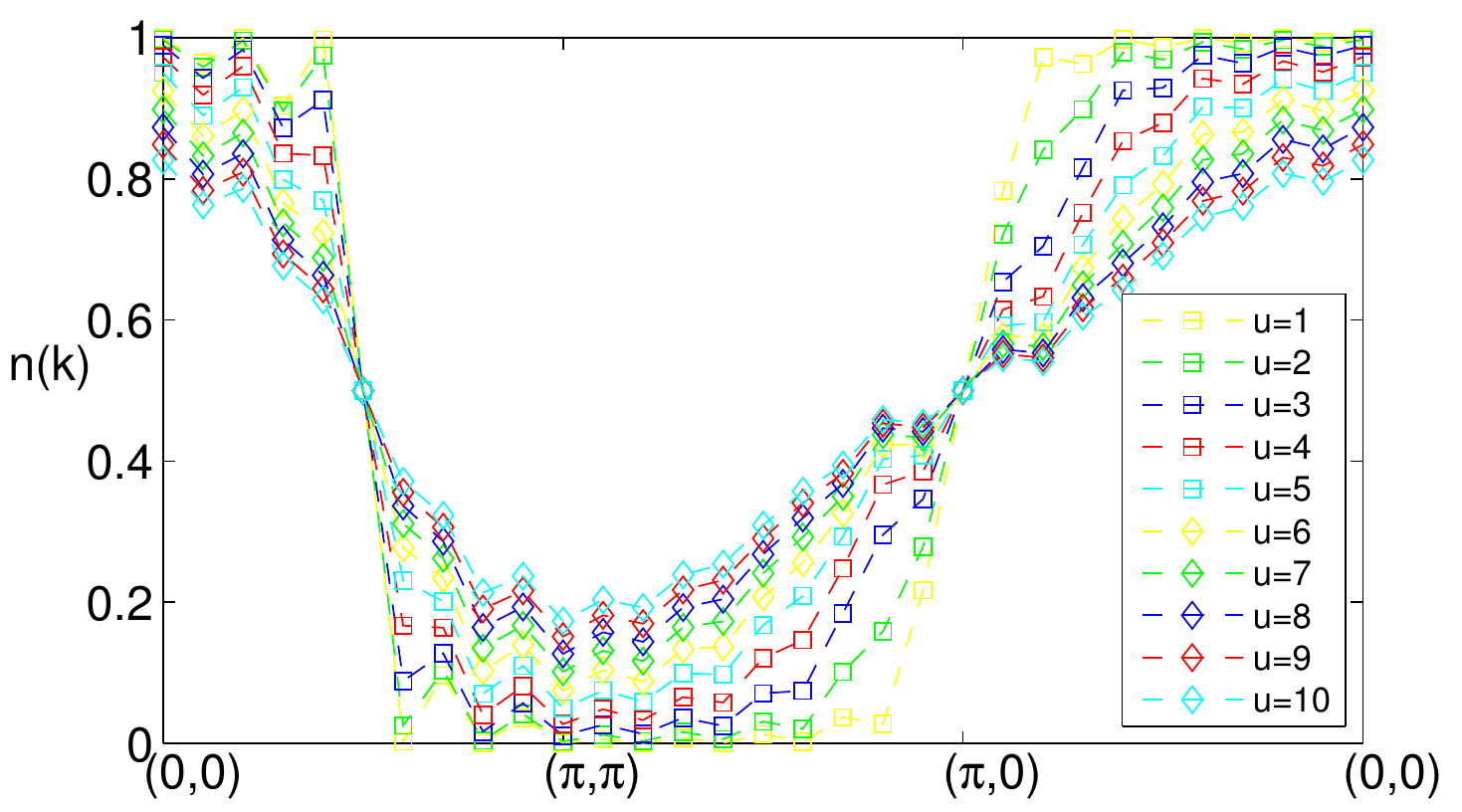}
\end{center}
\caption{Momentum distribution $n(k)$ of the ground state of the repulsive Hubbard model with an interaction ranging from $u=1,\ldots, 10$ at half-filling. While the Fermi surface is sharp for a weak coupling, we find a broadening with increasing $u$. (Compare with the QMC result, Fig. 1a of Ref.~\cite{Scarletter}.)\label{fig:Momentum_Distribution} }
\end{figure}

In the following, we consider two-particle properties of the system. To this end,  we study as in Ref.~\cite{Scarletter} the following magnetic properties of the system:  

\begin{itemize}
\item The equal-time spin-spin correlation function $C(\vec y) = \langle (n_{(\vec x + \vec y) \uparrow} - n_{(\vec x + \vec y) \downarrow})(n_{\vec x  \uparrow} - n_{\vec x \downarrow}) \rangle$.
\item The local moment $C(0,0) = \langle (n_{\vec x \uparrow} - n_{\vec x \downarrow})^2\rangle$.
\item The magnetic correlation function $S(k) = \sum_{\vec x}e^{i\vec k \cdot \vec x}C(\vec x)$.
\end{itemize}

\begin{figure}
\begin{center}
\includegraphics[width = 0.9\columnwidth]{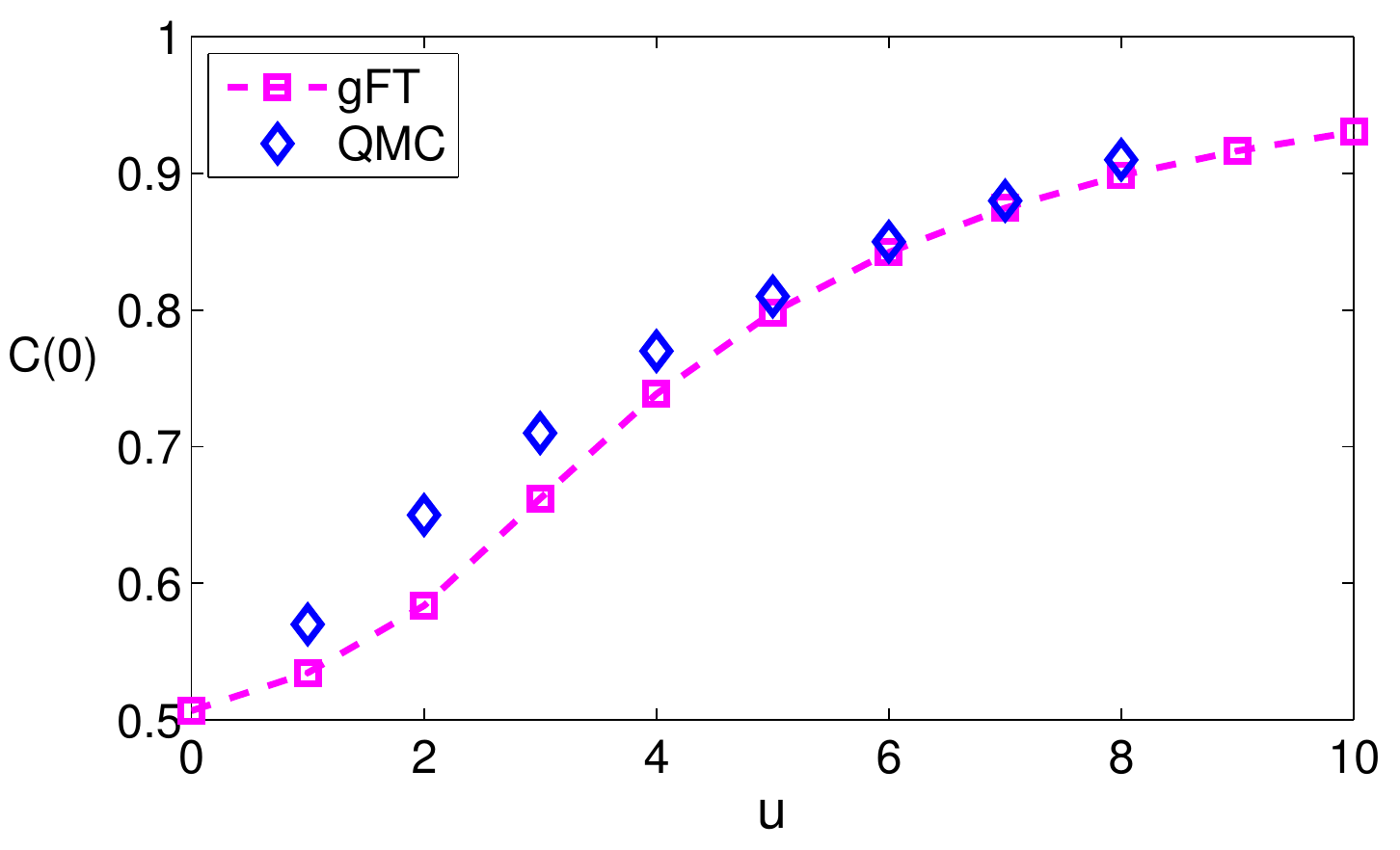}
\end{center}
\caption{Local moment $C(0,0) =  \langle (n_{\vec x \uparrow} - n_{\vec x \downarrow})^2\rangle$ of the ground state of the repulsive Hubbard model for an interaction $u = 0, \ldots, 10$ at half-filling. Starting from the non-interacting limit, where $C(0,0) = \tfrac{1}{2}$, the system approaches a value of $C(0,0) = 1$ with increasing interaction, corresponding to a phase with no double-occupancy. While the purple lines and squares depict the values of the magnetic moment obtained with our algorithm, the blue triangles represent the QMC results of Ref.~\cite{Scarletter}, Fig.~4. \label{fig:Spin_Spin_C0}}
\end{figure}

In Fig.~\ref{fig:Spin_Spin_C0} we have depicted the local magnetic moment of the ground state, $C(0,0) = \langle (n_{\vec x \uparrow} - n_{\vec x \downarrow})^2\rangle$, for an interaction strength $u = 0, \ldots, 10$ at half-filling. The purple squares depict the results obtained via our algorithm, while the blue diamonds represent the QMC results of ~\cite{Scarletter}. Singly occupied sites fulfill $C(0,0) = 1$, while for empty or doubly occupied sites $C(0,0)=0$. In the non-interacting regime, all four states (full, empty, single occupied in the two spin-states) are equally probable, so that $C(0,0) = \tfrac{1}{2}$. We find that the on-site repulsion suppresses doubly occupied, and for a half-filled lattice also the empty configuration. With increasing $u$ the singly occupied configuration is enhanced, leading to a well-formed moment at each site. 

\begin{figure}
\begin{center}
\includegraphics[width = 0.95\columnwidth]{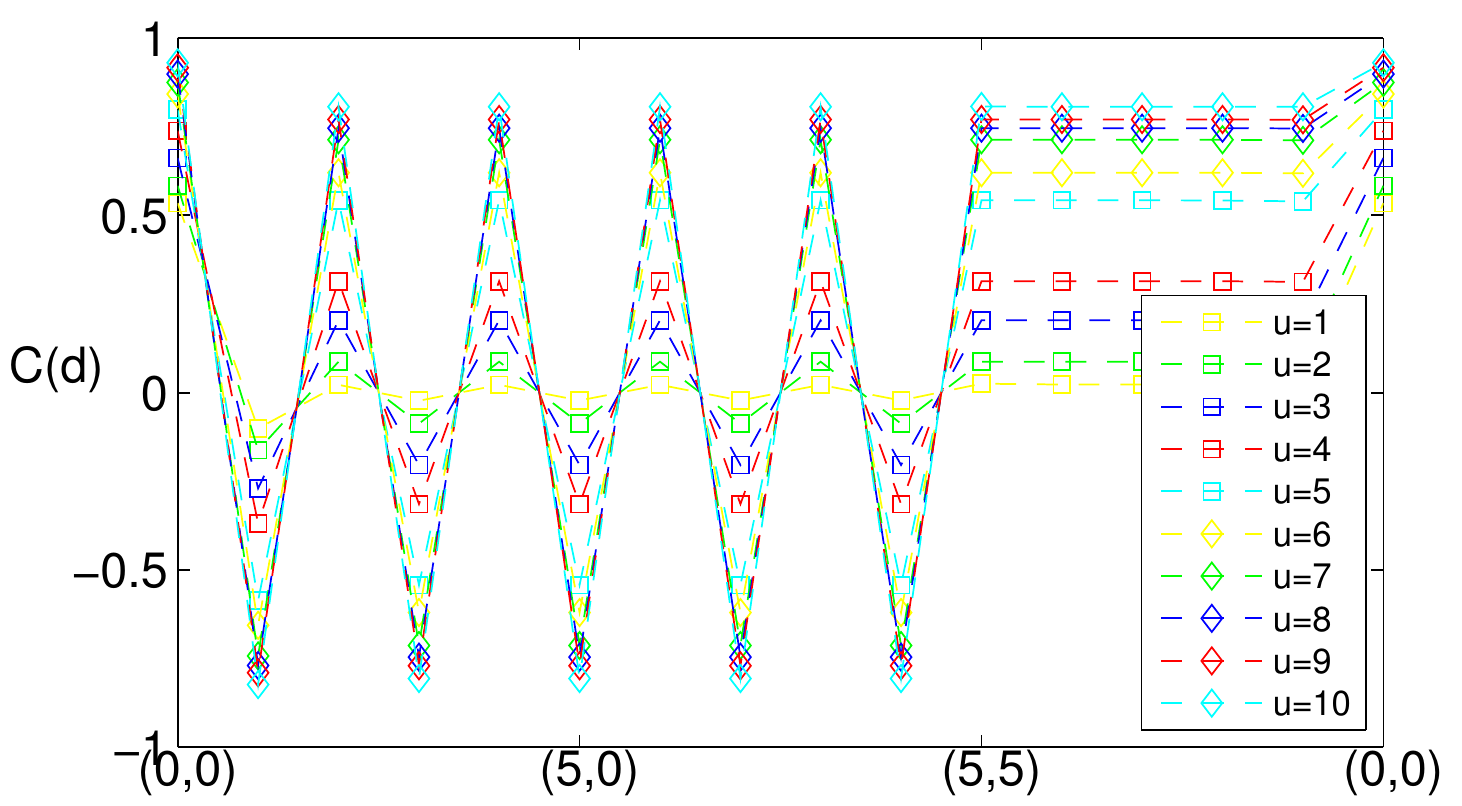}
\end{center}
\caption{Equal-time spin-spin correlation function $C(\vec y) = \langle (n_{(\vec x + \vec y) \uparrow} - n_{(\vec x + \vec y) \downarrow})(n_{\vec x  \uparrow} - n_{\vec x + \downarrow}) \rangle$ of the ground state for an interaction $u=1,\ldots 10$ and half-filling. Antiferromagnetic correlations increase with the interaction strength, but extend even for small interactions over the entire lattice. Compare with Ref.~\cite{Scarletter} Fig.~7, where $C(\vec y)$ has been obtained for a $24 \times 24$ lattice. \label{fig:Spin_Spin_Cd}}
\end{figure}

Next, we study the equal-time spin-spin correlation function $C(\vec y) = \langle (n_{(\vec x + \vec y) \uparrow} - n_{(\vec x + \vec y)\downarrow})(n_{\vec x \uparrow} - n_{\vec x \downarrow}) \rangle$. This quantity measures the extend to which two spins on site $\vec x$ and $\vec x+ \vec y$ are aligned. Note that the definition of $C(\vec y)$ is rotationally invariant. Our results are presented in Fig. ~\ref{fig:Spin_Spin_Cd}. The correlations extend, already for small values of $u$, over the entire lattice, as it is predicted for a half-filled lattice. As one would expect intuitively, the antiferromagnetic correlations are enhanced with growing interaction strength. Our results show the same qualitative behavior as in ~\cite{Scarletter}, where $C(\vec y)$ has been calculated for a $24 \times 24$ square lattices.

\begin{figure}
\begin{center}
\includegraphics[width = 0.95\columnwidth]{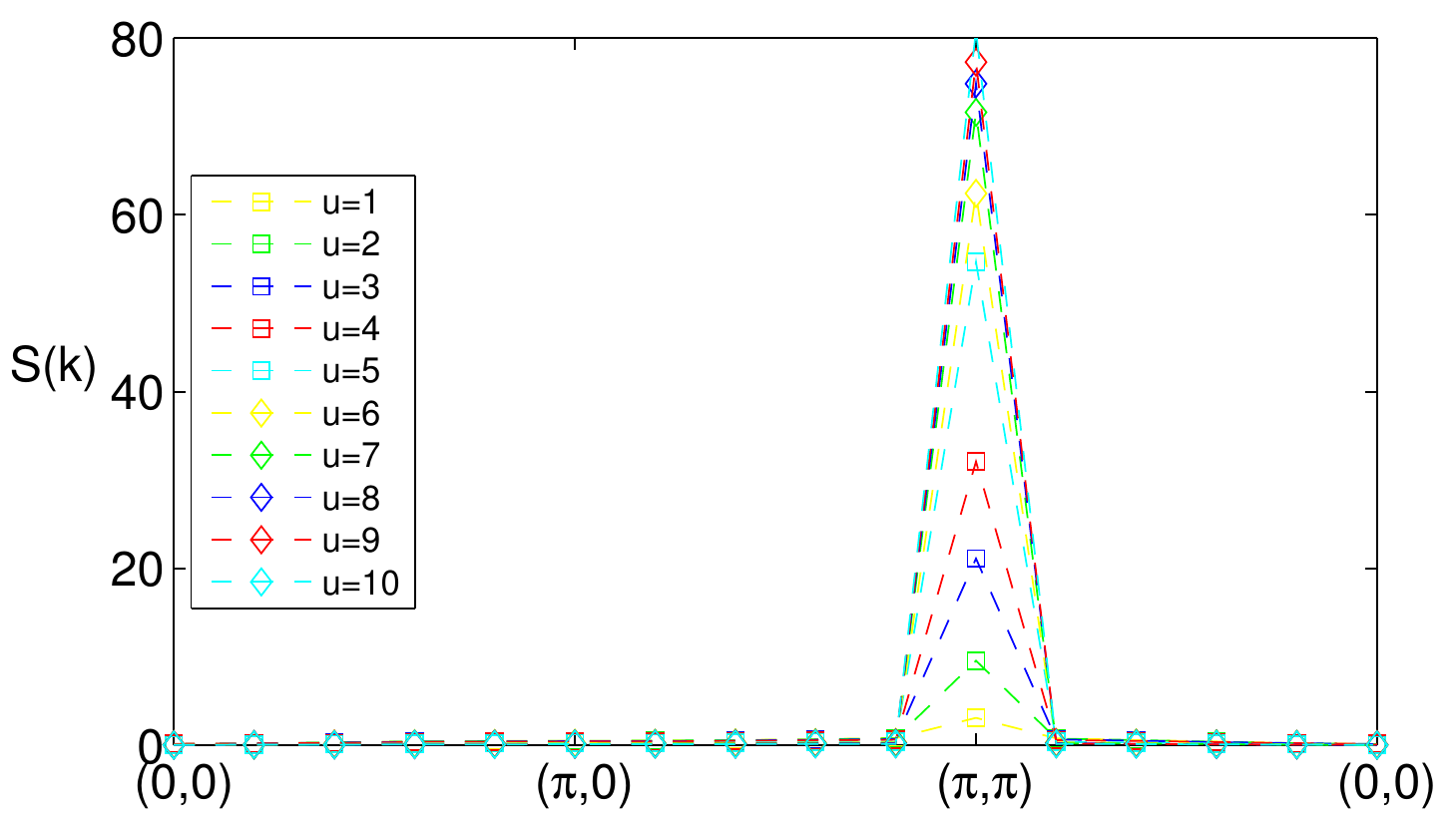}
\end{center}
\caption{Magnetic correlation function $S(k) = \sum_{\vec l}e^{i\vec k \cdot \vec l}C(\vec l)$ for the ground state for $u=1, \ldots, 10$ at half-filling. The sharp peak at $(\pi,\pi)$ emphasizes the antiferromagnetic correlations inherent in a half-filled lattice. (Compare with Fig. 8 of Ref.~\cite{Scarletter})\label{fig:Magnetic_Structure}}
\end{figure}

Finally, we consider the magnetic correlation function $S(k) = \sum_{\vec l}e^{i\vec k \cdot \vec l}C(\vec l)$ in Fig.~\ref{fig:Magnetic_Structure}. A sharp peak occurs at momentum $(\pi, \pi)$, emphasizing the antiferromagnetic correlations. (Compare with Fig. 8 of Ref.~\cite{Scarletter}, where $S(k)$ is depicted for $u=2$). Further, we see that the magnetic correlations depend strongly on the interaction strength.

\subsubsection{Ground state properties of a doped system}
\begin{figure}
\begin{center}
\includegraphics[width=0.95\columnwidth]{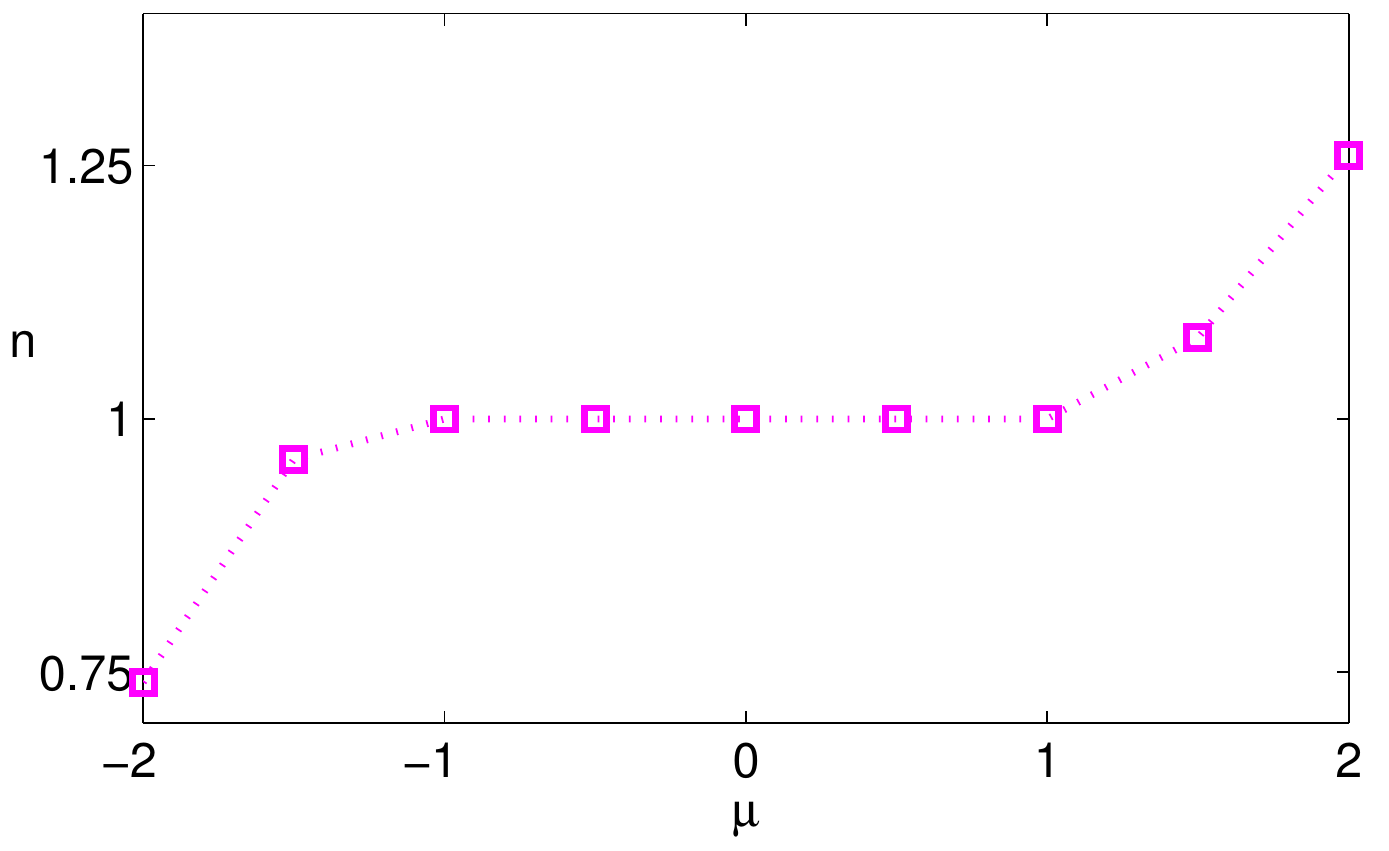}
\end{center}
\caption{Local density $n_x = n_{x\uparrow} + n_{x \downarrow}$ as a function of the chemical potential for the ground state at $u=6$. (Compare with Fig. 2a of Ref.~\cite{trivedi-2009}) \label{fig:Density_mu}}
\end{figure}

Next, we consider the case of a doped system, i.e. $\mu \neq 0$. This instance has been studied e.g. in Ref.~\cite{trivedi-2009}. There, the local density $n_x$ was calculated as a function of the chemical potential. The results obtained via our approach (see Fig. \ref{fig:Density_mu}) show a qualitative agreement with Fig. 2a of Ref.~\cite{trivedi-2009}, where a density of $n(\mu) = 1$ is found for $\mu \approx -1, \ldots, 1$. We find that for $\mu = 0$ we have half-filling, as expected, while an increase in the chemical potential results finally in doubly occupied sites.

We include here a remark on the limitations of gHFT. It is believed that the doped Hubbard model with repulsive interaction is a highly correlated state, exhibiting pairing for some value of the doping. However, for $u \in [0,10]$ and $\mu \in [-5,5]$ we only find an unpaired phase. This is in agreement with Thm. 2.11 of Ref. \cite{bach-1993}, where it has been proven that for any positive definite potential $U$ the gHF ground state is unpaired, independent of the doping.
\subsubsection{Thermal state properties for half-filling}
\begin{figure}[t]
\begin{center}
\includegraphics[width = 0.85\columnwidth]{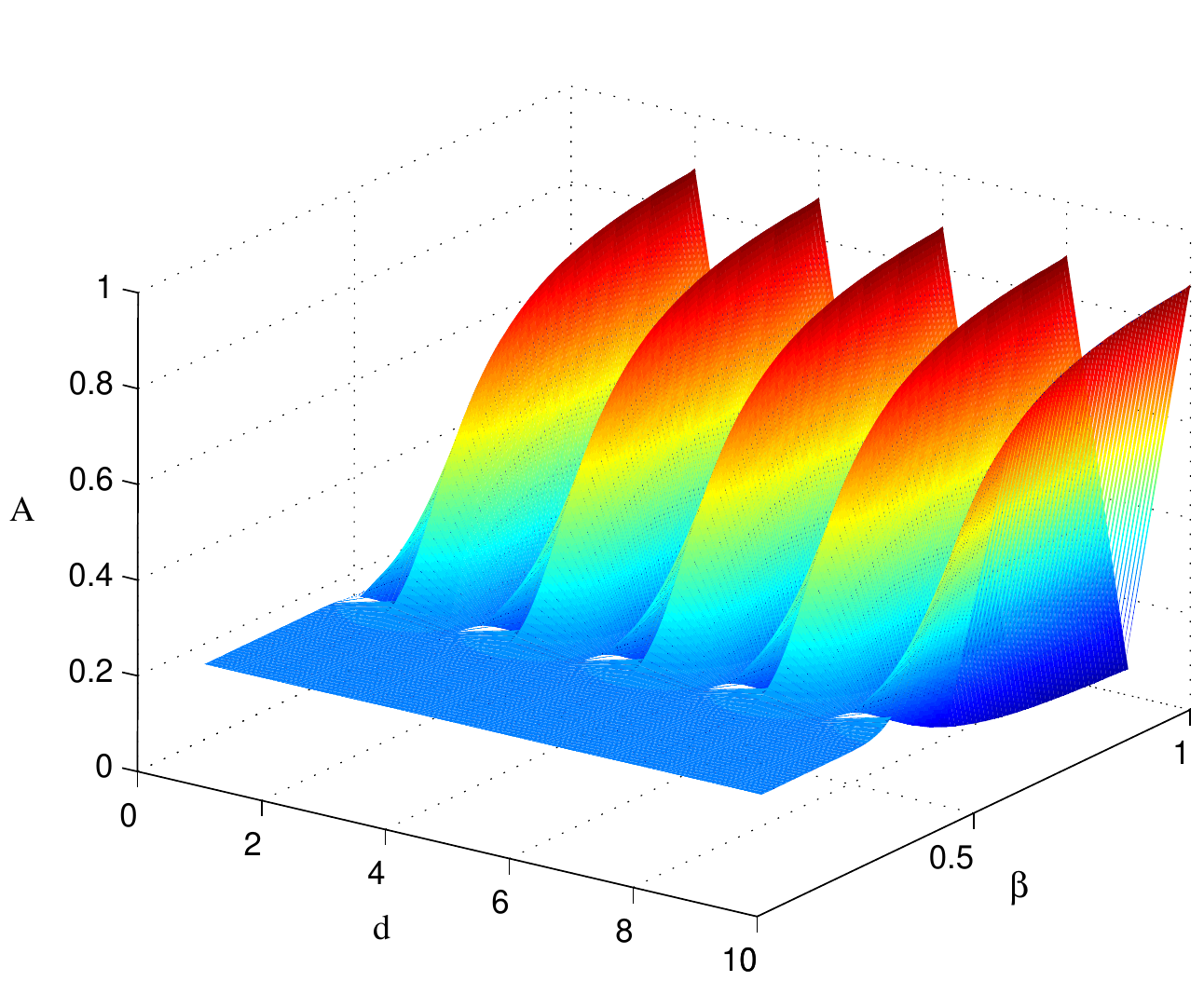}
\end{center}
\caption{Antiferromagnetic correlation $A = \langle n_{x\uparrow}n_{x+d \downarrow}\rangle$ in for the repulsive Hubbard model at half-filling as a
function of the inverse temperature $\beta \in [0,1]$ for an interaction strength $u=10$ and lattice spacing $d=1, \ldots, 10$. We find that the antiferromagnetic order is destroyed with increasing temperature due to thermal fluctuations. \label{fig:antiferro}}
\end{figure}
In this Subsection we consider some finite temperature properties of the repulsive Hubbard model, and concentrate again on the case of half-filling. We have seen in the last Section that gHFT provides us with a powerful tool to study the magnetic properties of fermions with repulsive interactions in a lattice. Especially, we have seen that the ground state of the system shows antiferromagnetic correlations, already for small values of the interaction strength. These are expected to decrease with increasing temperature, since the thermal fluctuations destroy the antiferromagnetic order. We find indeed this behavior, which is depicted in Fig.~\ref{fig:antiferro}. There, we have plotted the order parameter 
\begin{equation}
A(d) = \langle n_{x\uparrow}n_{x+d \downarrow}\rangle \label{eq:Def_AF}
\end{equation}
as a function of the  of the inverse temperature for $u=10$. We have also derived the critical exponent $\alpha$ given by $A(T)\sim (T-T_c)^{\alpha}$ for lattice sizes $8\times 8$ and $10\times 10$, leading to $\alpha \approx 1.3-1.4$. 

\begin{figure}[t]
\begin{center}
\includegraphics[width=0.95\columnwidth]{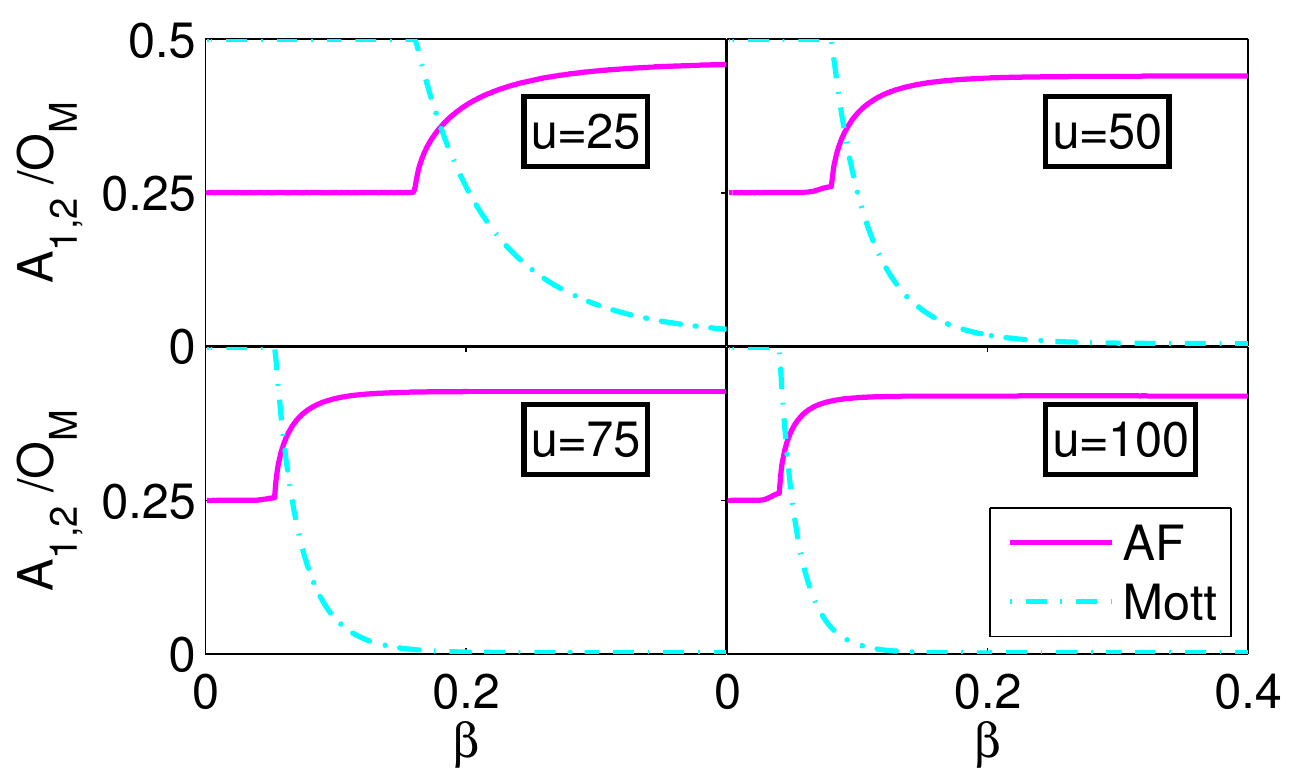}
\end{center}
\caption{Value of the antiferromagnetic order parameter $A_{1,2}=A(1)$ (c.f. Eq. \eqref{eq:Def_AF}) of nearest-neighbors (solid purple line) and order parameter for the Mott phase, $O_M = \langle (n_x)^2 \rangle- \langle n_x\rangle^2$ (dash-dotted blue line), for the attractive Hubbard model with interaction strength $u=25, 50, 75, 100$ as a function of the inverse temperature, $\beta \in [0, 0.4]$. We find an absence of the Mott-gap.   
 \label{fig:Mott_AF}}
\end{figure}

The antiferromagnetic phase is further characterized by the fact that the local fluctuations in the particle number vanish, i.e. we are in a Mott phase. This phase is predicted to last for even higher temperatures than the antiferromagnetic phase (''Mott-gap''), since additional energy is needed to delocalize the fermions~\cite{ScalapinoMottGap}. We investigate this behavior in Fig. \ref{fig:Mott_AF}. There, the solid purple line shows the value of $A_{1,2}=A(1)$ (c.f. Eq. \eqref{eq:Def_AF}), the antiferromagnetic correlations of nearest-neighbors, while the blue dotted line is the order parameter for the Mott phase, $O_M = \langle (n_x)^2 \rangle- \langle n_x\rangle^2$. We see that independent of the interaction strength the Mott phase as well as the antiferromagnetic phase break down at the same temperature, i.e. we do not find a Mott-gap. This observation can be explained by the fact that the FGS cannot represent the Mott phase, as we have already stated in the introduction. 

The fact that there exists no Gaussian state corresponding to the Mott phase except the state of the antiferromagentic phase also leads to an incorrect transition temperature from the antiferromagnetic to the normal phase. For the exact solution of the Hubbard model the antiferomagnetic correlations are predicted to be destroyed at a temperature $T_c \sim t^2/u$ which is the energy scale of the superexchange energy.  However, in our approach the antiferromagnetic correlations appear as soon as we find one particle per site, leading to a transition temperature $T_c \sim u$ (cf. Fig~\ref{fig:Mott_AF}).

\section{The Hubbard model in an external trap}\label{sec:Hubbard_trap}
\begin{figure}[t]
\begin{center}
\includegraphics[width = \columnwidth]{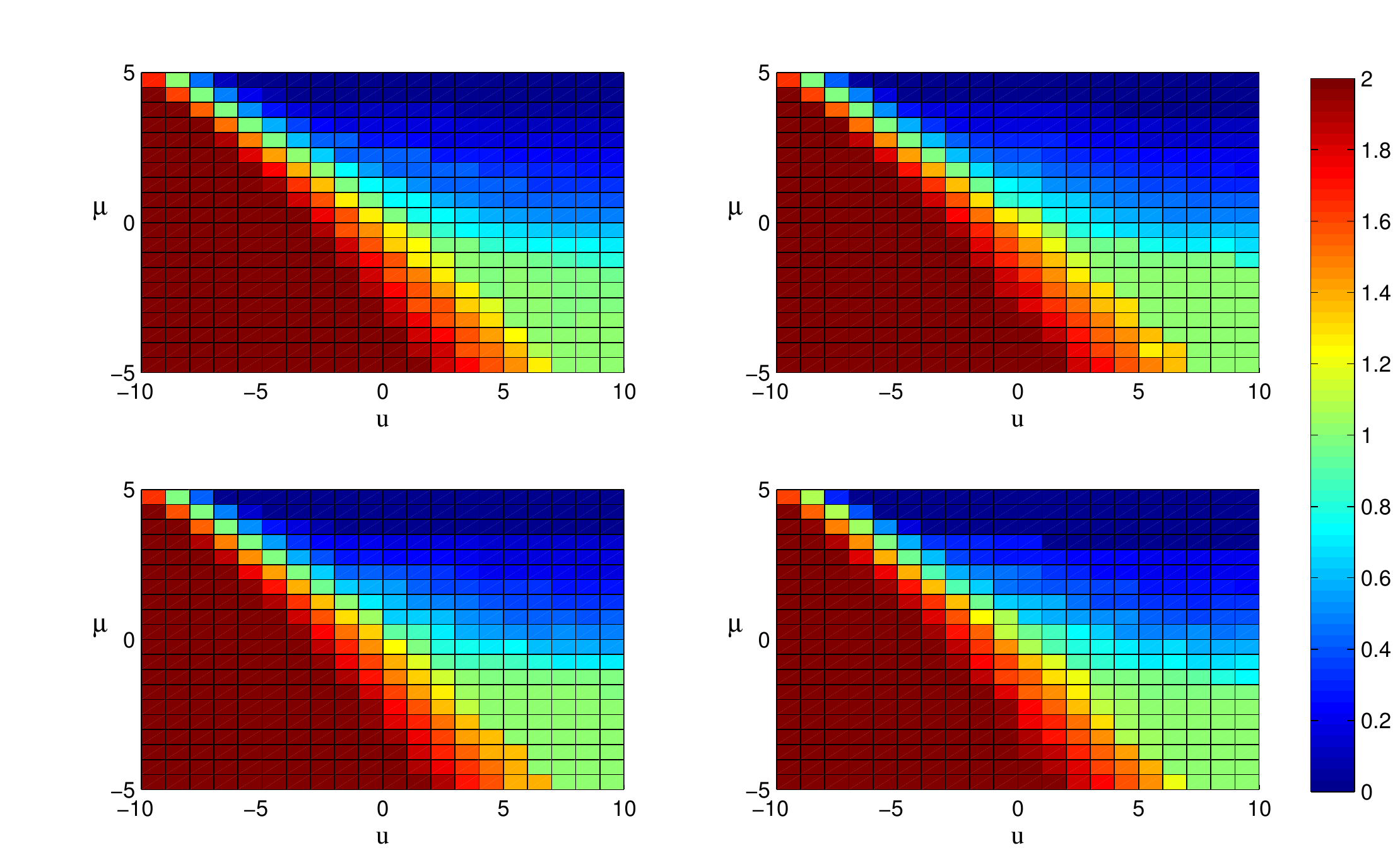}
\put(-220,141){\scriptsize   a) $V_t = 0$ }
\put(-110,141){\scriptsize  b) $V_t = 0.05$ }
\put(-220,67){\scriptsize c) $V_t = 0.1$ }
\put(-110,67){\scriptsize  d) $V_t = 0.25$ }
\end{center}
\caption{Density at the center of the trap for the ground state  as a function of the chemical potential $\mu$ and the interaction strength $u$ for three different trapping potentials $V_t = 0.05, 0.1, 0.25$ (b) - (d), as well as for the translationally invariant case (a).
\label{fig:Density_trap_mu_u}}
\end{figure}
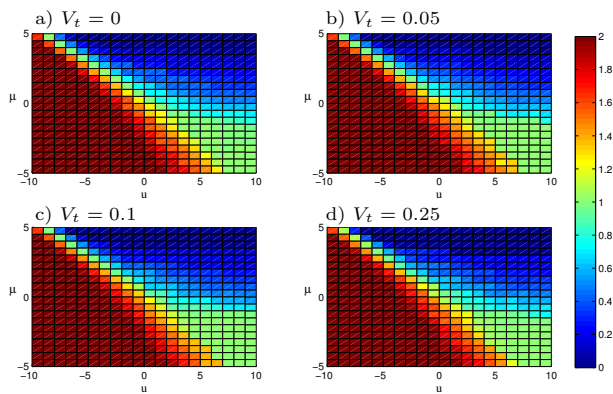

While the translationally invariant Hubbard model allows for an elegant analytic solution within generalized Hartree Fock theory, experimental setups break this symmetry. Since the particles have to be spatially confined, an external trapping potential is needed. In many setups, this trapping is realized via a harmonic confinement. The Hamiltonian of the system is then altered by a position-dependent chemical potential, i.e.
\begin{multline}\label{eq:Hubbard_H_2d_trap}
H= t\sum_{ \langle x,y  \rangle \in \Lambda, \sigma}\ad_{x,\sigma}a_{y,\sigma} \\+ u \sum_{i\in
\Lambda}n_{x\uparrow}n_{x\downarrow} +
\sum_{x,\sigma}\mu_{x} n_{x,\sigma},
\end{multline}
where for $x = x(h,v)$ with coordinates $(h,v)$ we have
\begin{equation}
\mu_{x} = \mu + V_t \left[\left(\frac{N_h+1}{2}- h \right)^2  + \left(\frac{N_v+1}{2} - v\right)^2
\right].
\end{equation}
In the following we investigate how the external confinement alters the physical properties of the attractive Hubbard model (with open boundary conditions) at zero and finite temperature. To gain first insights into the behavior of the system, we have depicted in Fig.~\ref{fig:Density_trap_mu_u} the density at the center of the trap as a function of the interaction strength and the chemical potential for three different trapping potentials (Fig.~\ref{fig:Density_trap_mu_u} b-d), as well as for the translationally invariant case (Fig.~\ref{fig:Density_trap_mu_u} a). We see that the filling in the center of the trap is only altered slightly be the external potential. In the following, we call a system half-filled if there is one particle at the center of the trap. We find that the condition for half-filling does not depend much on the trapping potential, and we can use the results from Fig.~\ref{fig:Density_trap_mu_u} to adjust the chemical potential in order to obtain half-filling in the following.

\subsection{Attractive Hubbard model in a trap}
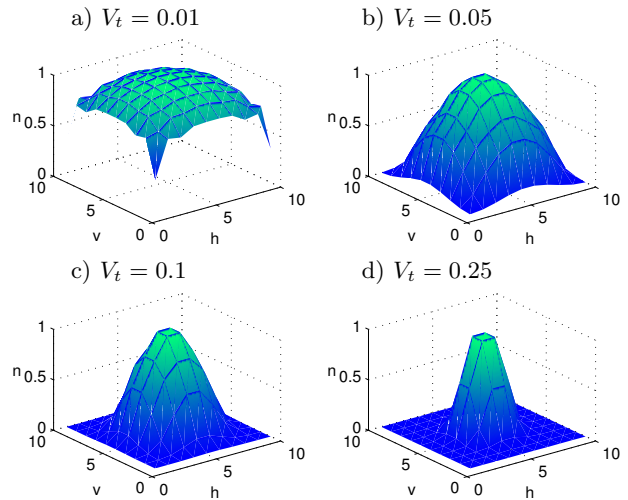
\begin{figure}[tbh]
\begin{center}
\includegraphics[width = 0.95\columnwidth]{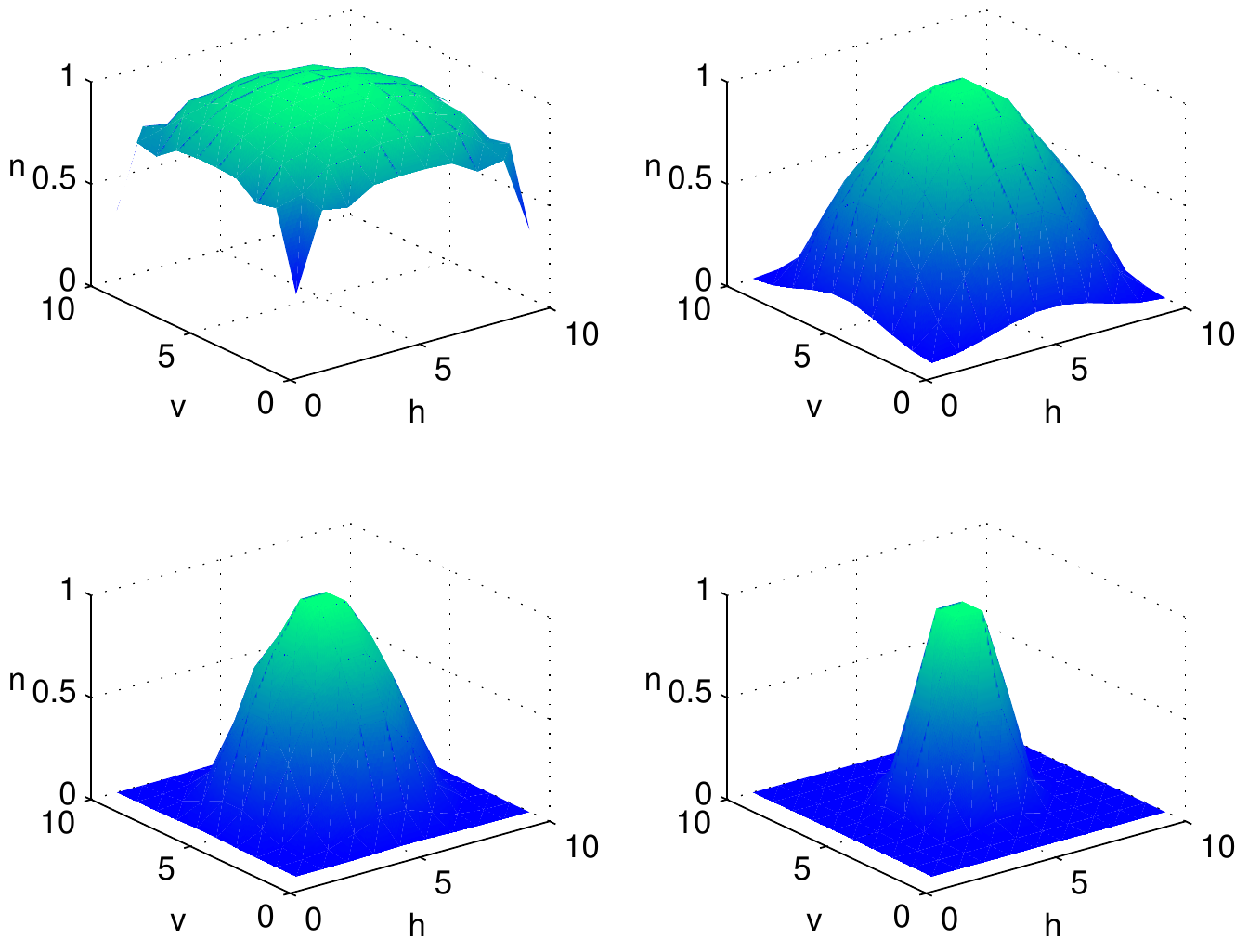}
\put(-210,182){a) $V_t = 0.01$ }
\put(-100,182){b) $V_t = 0.05$ }
\put(-210,86){c) $V_t = 0.1$ }
\put(-100,86){d) $V_t = 0.25$ }
\end{center}
\caption{Density profiles for the ground state ( $T = 0$) for the four different trapping potentials $V_t = 0.01, 0.05, 0.1, 0.25$ at half-filling (i.e. one particle at the center of the trap) and $u = -5$.\label{fig:density_trap_uminus5}}
\end{figure}

We start with a closer investigation of the effects of an external trapping potential on the physics of the attractive Hubbard model. In Fig.~\ref{fig:density_trap_uminus5} we present the density distribution for $u=-5$ at half-filling for four different trapping potentials $V_t = 0.01, 0.05, 0.1, 0.25$. We see that for $V_t = 0.01$ the density distribution is close to that of a free system with open boundary conditions, so that  boundary effects are expected in this case. Hence, we will consider only $V_t =0.05, 0.1, 0.25$ from now on, and compare the results to the translationally invariant model with periodic boundary conditions.

\begin{figure}[t]
\begin{center}
\includegraphics[width = 0.95\columnwidth]{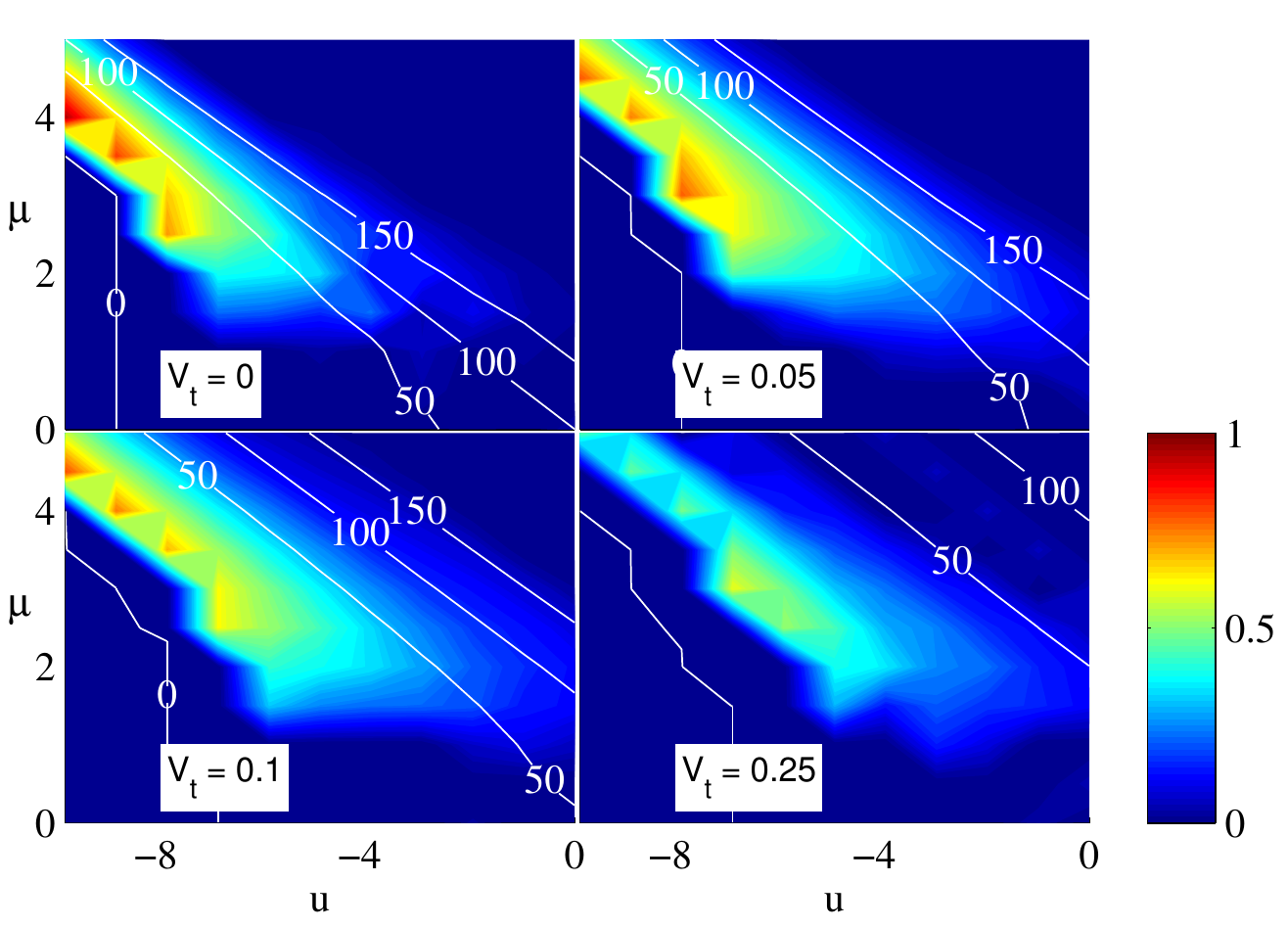}
\end{center}
\caption{Pairing of the ground state  ( $T = 0$) of the attractive Hubbard model with an external confinement $V_t =  0.05, 0.1, 0.25$, as well as for the translationally invariant case ($V_t = 0$), for an interaction range of $u \in [-10,0]$ and for a chemical potential $\mu \in [0,5]$.
\label{fig:Pairing_trap}}
\end{figure}

Now we investigate how the phase diagram of the pairing changes under the influence of an external confinement. In Fig.\ref{fig:Pairing_trap} we have depicted how the pairing per particle as a function of the interaction strength $u$ and the chemical potential $\mu$ changes under the influence of an external trap. We have added lines for $50$, $100$ and $150$ particles as a guide to the eye. 

\begin{figure}[t]
\begin{center}
\includegraphics[width = 0.95\columnwidth]{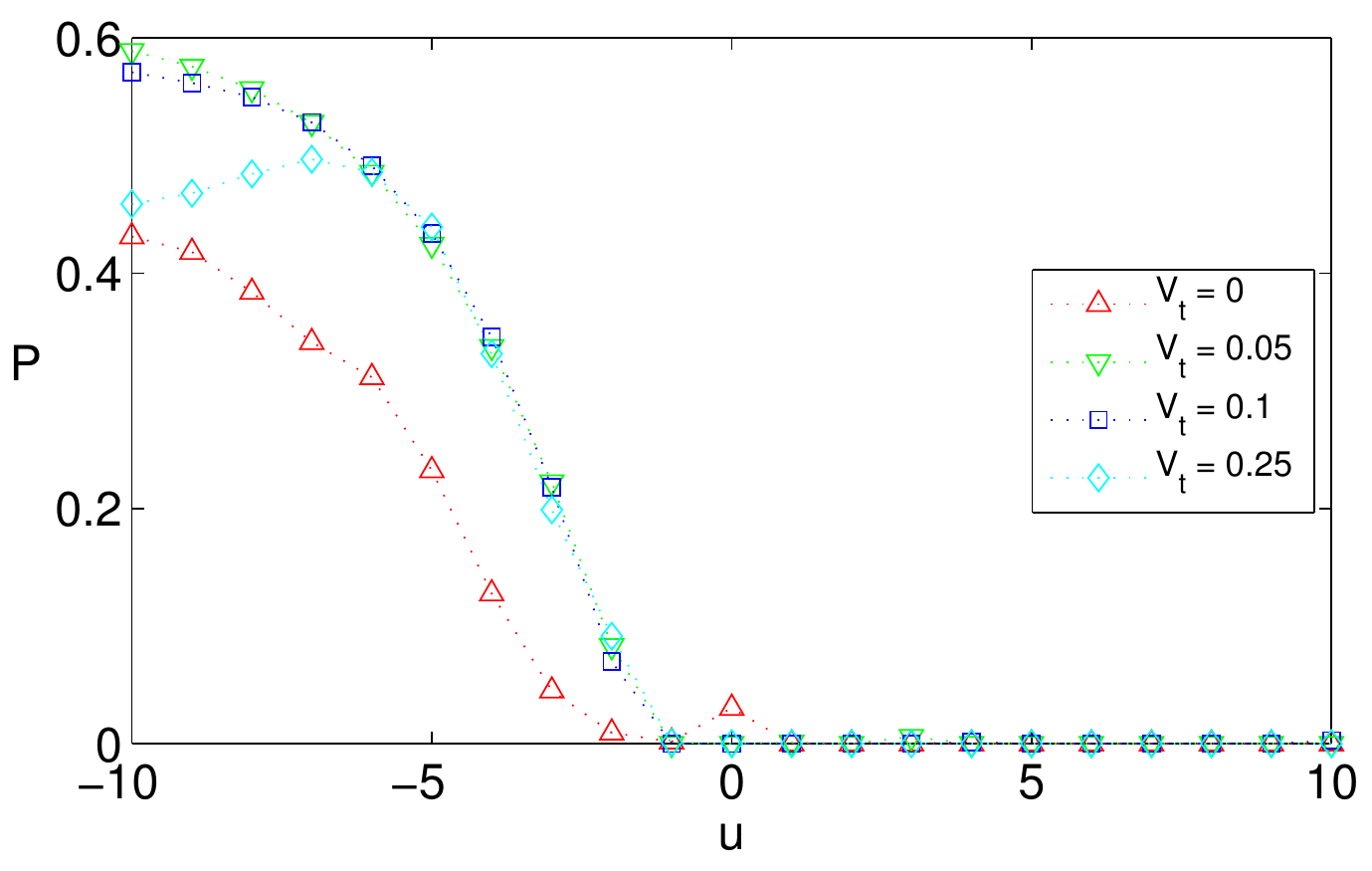}
\end{center}
\caption{Pairing per particle for the ground state ($T = 0$) for the attractive Hubbard model with an external confinement $V_t = 0.05, 0.1, 0.25$ at half-filling, as well as for the translationally invariant case ($V_t = 0$). (Note that in that case the absolute value of the pairing is not uniquely determined due to an additional Gauge freedom -- see text). If the external trapping potential becomes too strong, the paired phase is suppressed.
\label{fig:Pairing_trap_T0}}
\end{figure}

To gain further insight into the influence of the trap, we have depicted in Fig.~\ref{fig:Pairing_trap_T0} the pairing per particle for a half-filled trap, i.e. a setting where we find one particle at the center of the trap, and compare with the translationally invariant case. Note that for a translationally invariant system an additional Gauge symmetry emerges at half-filling  ~\cite{bach-1993}, and the underlying Gauge transformation does not leave the pairing invariant. However, this Gauge symmetry is broken for an inhomogeneous system, and we can only perform a qualitative comparison of the translationally invariant and trapped system. In this respect, we observe that the pairing decreases with decreasing interaction strength, both in the translationally invariant and in the trapped system, as expected. Further, we see that a very strong confinement can lead to a slight decrease in the pairing per particle. This effect can be understood in the following way: Since we fix the number of particles at the center of the trap to one, and the external trap acts like a position-dependent chemical potential, a strong trap will significantly reduce the number of particles at the boundary of the lattice, and the paired phase is suppressed.

\begin{figure}[t]
\begin{center}
\includegraphics[width = 0.8\columnwidth]{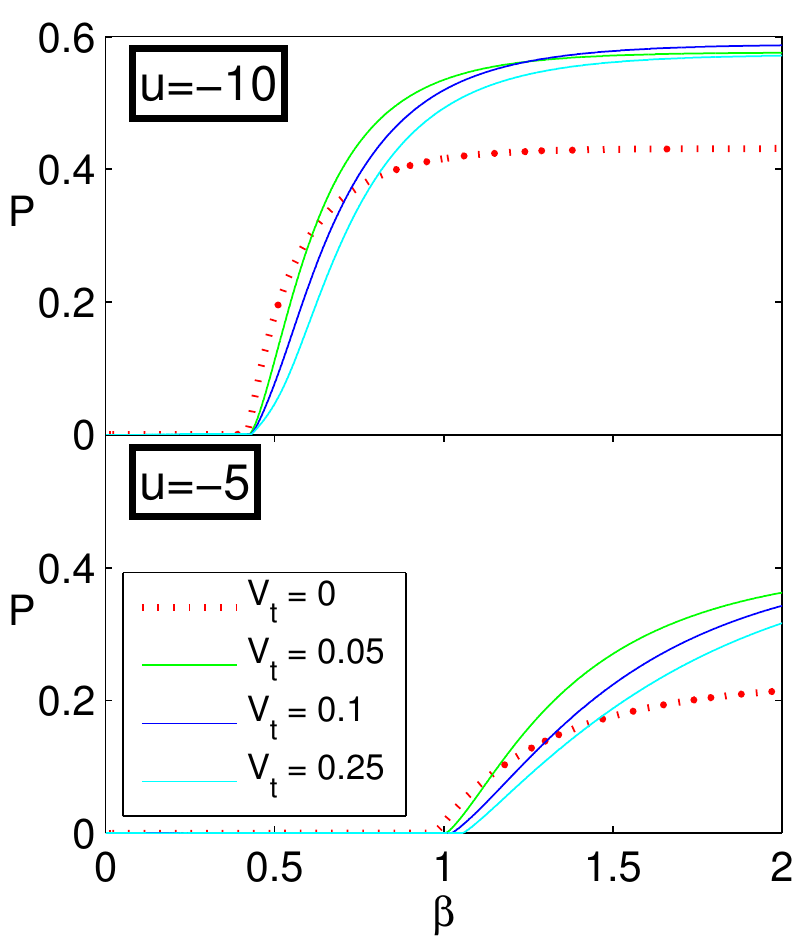}
\end{center}
\caption{Influence of an external trap on the pairing  as a function of the temperature for  $u =-10$ and $u=-5$ and trapping potentials $V_t =0.05, 0.1, 0.25$, as well as for the translationally invariant case ($V_t = 0$) at half-filling. We find that the external confinement has no influence on the temperature $\beta_c$ where the transition from the paired to the unpaired phase occurs. Note that if the temperature is high enough, the gHF Gibbs state becomes unique for the translationally invariant case (see text).\label{fig:Pairing_finite_T_trap}}
\end{figure}

We continue our investigations on the influence of the external trapping potential by analyzing the temperature-dependence of the pairing in an external trap. In Fig.~\ref{fig:Pairing_finite_T_trap} we consider interactions $u =-10$ and $-5$ and a temperature regime of $\beta \in [0,2]$. First, we observe that the external trapping alters only slightly the temperature-dependence of the pairing. Further, the value of $\beta_c$, where a phase transition from a paired to a normal state occurs is not altered by the external confinement, and is identical to the translationally invariant case. At this point we would also like to mention that the Gauge freedom of the ground state is lifted with increasing temperature, and a unique gHF ground state is expected if the temperature becomes high enough ~\cite{bach-1993}. With the results depicted in Fig.~\ref{fig:Pairing_finite_T_trap} we can now conclude that for high enough temperature the pairing as well as the transition from a paired to an unpaired phase in a trap are also approximated well quantitatively  by a translationally invariant system.

\begin{figure}[tbh]
\begin{center}
\includegraphics[width = 0.95\columnwidth]{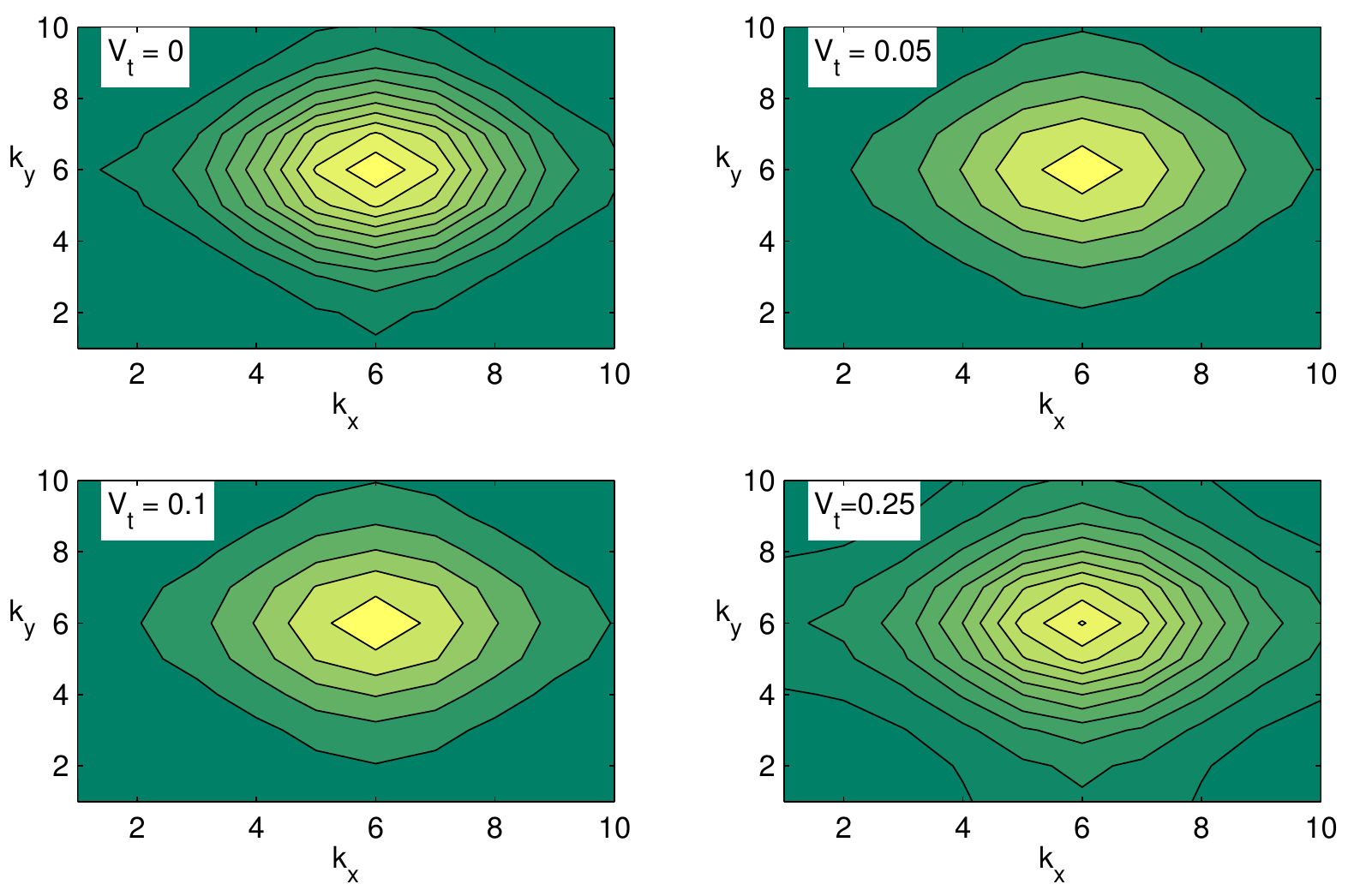}
\end{center}
\caption{Pair wave function in momentum space, $|\phi(k_1, k_2)|^ 2$ for the translationally invariant $(V_t = 0)$ and trapped $(V_t = 0.05, 0,1, 0.25)$ system for $u=-6$ at half-filling. The wave-function is always spherically symmetric, in agreement with s--wave superconductivity. \label{fig:Symmetry}}
\end{figure}

Finally, we address another question related to superconductivity, namely the symmetry of the pair wave function, i.e. the question if we have an s-- or p--wave superconductor. Every pure Gaussian state has a wave function of the form $|\Psi \rangle = \exp \left[\sum_{\vec x,\vec y} \phi_{\sigma, \sigma'}(\vec x, \vec y) \ad_{\vec x, \sigma} \ad_{\vec y,\sigma'}\right] \vac$. Since we consider a translationally invariant system, we have $ \phi_{\sigma, \sigma'}(\vec x, \vec y)  =  \phi_{\sigma, \sigma'}(\vec r) $, where $\vec r = \vec x - \vec y$. In Fig.~\ref{fig:Symmetry}, we have depicted the pair wave function in momentum space, $|\phi_{\uparrow, \downarrow}(k_1, k_2)|^ 2$  for a translationally invariant and trapped system for $u=-6$ and half-filling. We find for all trapping potentials a spherically symmetric wave function in agreement with s--wave superconductivity.

In summary, we find that a weak external confinement alters only slightly the physics of ground and thermal state of the attractive Hubbard model, so that experiments that require only a weak confinement will be well suited to gain further understanding of the $2d$ Hubbard model.

\section{Dynamical properties of the Hubbard model}\label{sec:Dynamics}
Current experimental setups with ultra cold gases in optical lattices have opened the door to an investigation of the dynamical behavior of fermionic lattice systems.  To this end, it has become possible to study processes like the BEC--BCS-crossover or dynamical quenches in the laboratory. Unfortunately, existing numerical techniques, like QMC, are not capable of describing the physics of time-evolution, so that numerical benchmark results for these processes are still missing so far. 

In this Section we apply the machinery of time-evolution within gHFT derived in Sec.~\ref{sec:real_time} to dynamical processes as they can be realized by current experimental setups. In the following we consider both, the case of a translationally invariant system with periodic boundary conditions as well as the case of an external trapping potential.

\subsection{Dynamical process of the translationally invariant Hubbard model}
In this Section we want to learn more about dynamical processes of the translationally invariant Hubbard model with periodic boundary conditions. To this end we have investigated a linear ramp of the interaction strength $u$ for various ramping times $T_f$, addressing the question if we can achieve an adiabatic evolution of the system. As an example, we have started from the ground state for $u=-7$ and have increased $u$ linearly to the final value $u = 7$. We have used ramping times $T_f = 1, 10, 100, 200, 400$ and have compared the energy of the evolving system, $E(u, T_f)$, with its ground state energy, $E_0(u)$, for a given $u$. 

\begin{figure}[t]
\begin{center}
\includegraphics[width = 0.95\columnwidth]{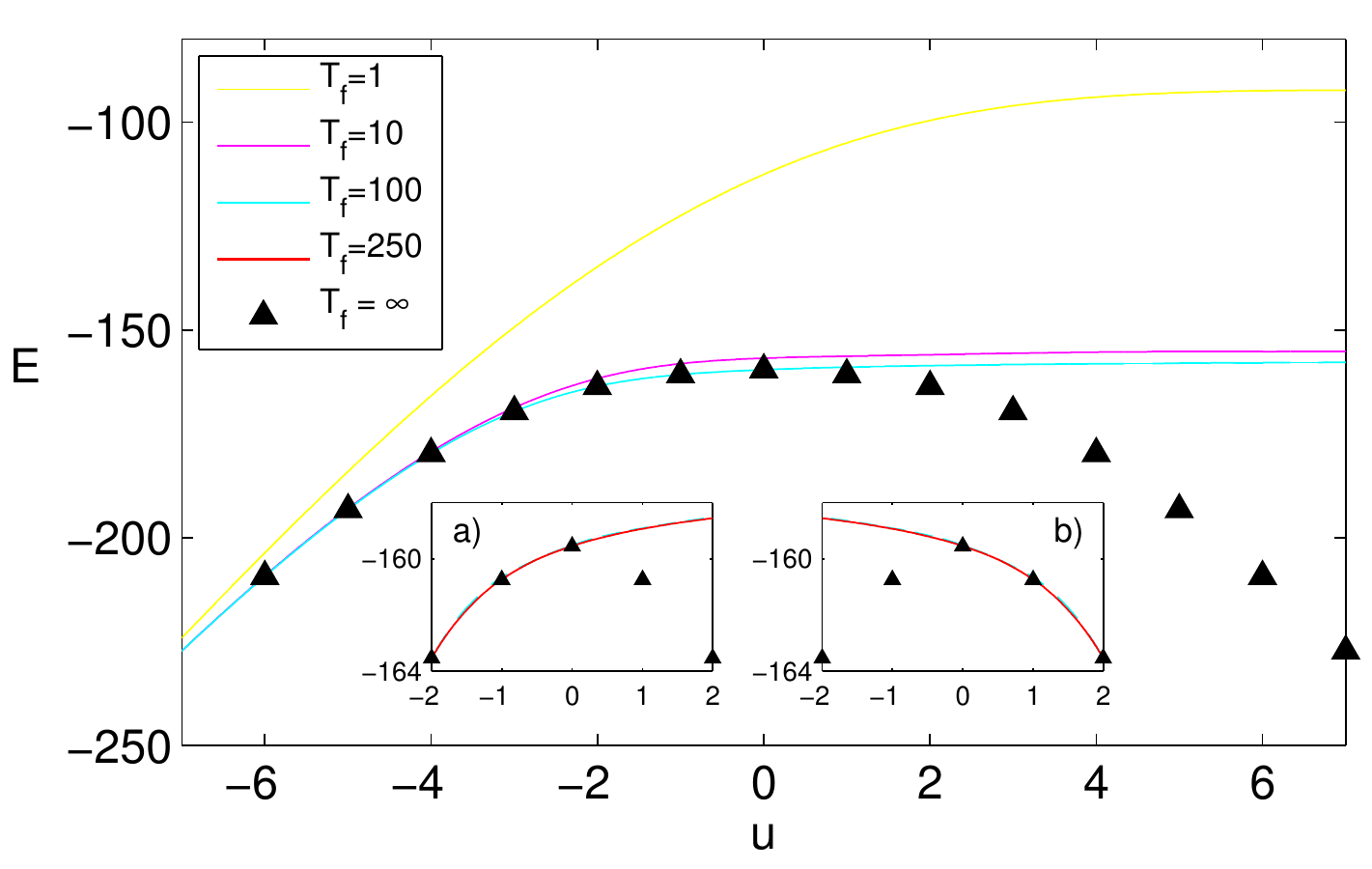}
\end{center}
\caption{BEC--BCS-crossover for the translationally invariant Hubbard model at half-filling. Starting from the ground state for $u=-7$ we perform a linear ramp to the final value $u=7$ with ramping times $T_f = 1, 10, 100$ and compare the energy during the evolution with the ground state energy for various $u$ (black triangles). The evolution is adiabatic for all negative $u$ when $T_f = 100$, while adiabaticity is no longer given for $u$ positive. This problem cannot be resolved by longer ramping times (a).  Further, starting from $u>0$ and ramping to $u<0$ leads to the same observation: As soon as $u=0$ is reached the evolution of the system does not follow the ground state energy any more (b).  \label{fig:Crossover_energy_inset}}
\end{figure}

As a result, we observe first that $E(u, T_f)$ has the same form for $T_f = 100, 200, 400$, so that we have only depicted the evolutions for $T_f  = 1, 10, 100$ in Fig.~\ref{fig:Crossover_energy_inset}. We find that for $T_f=100$ the evolution is adiabatic for all negative $u$. However, from $u=0$ onwards, the energy starts to deviate more and more from the true ground state energy. An increase in ramping time does not help to resolve this problem (Fig.~\ref{fig:Crossover_energy_inset}a). In  Fig.~\ref{fig:Crossover_energy_inset}b we have depicted a ramp from $u=2$ to $u=-2$, and also find that adiabaticity is destroyed when we cross the value $u=0$.

\begin{figure}[t]
\begin{center}
\includegraphics[width = 0.95\columnwidth]{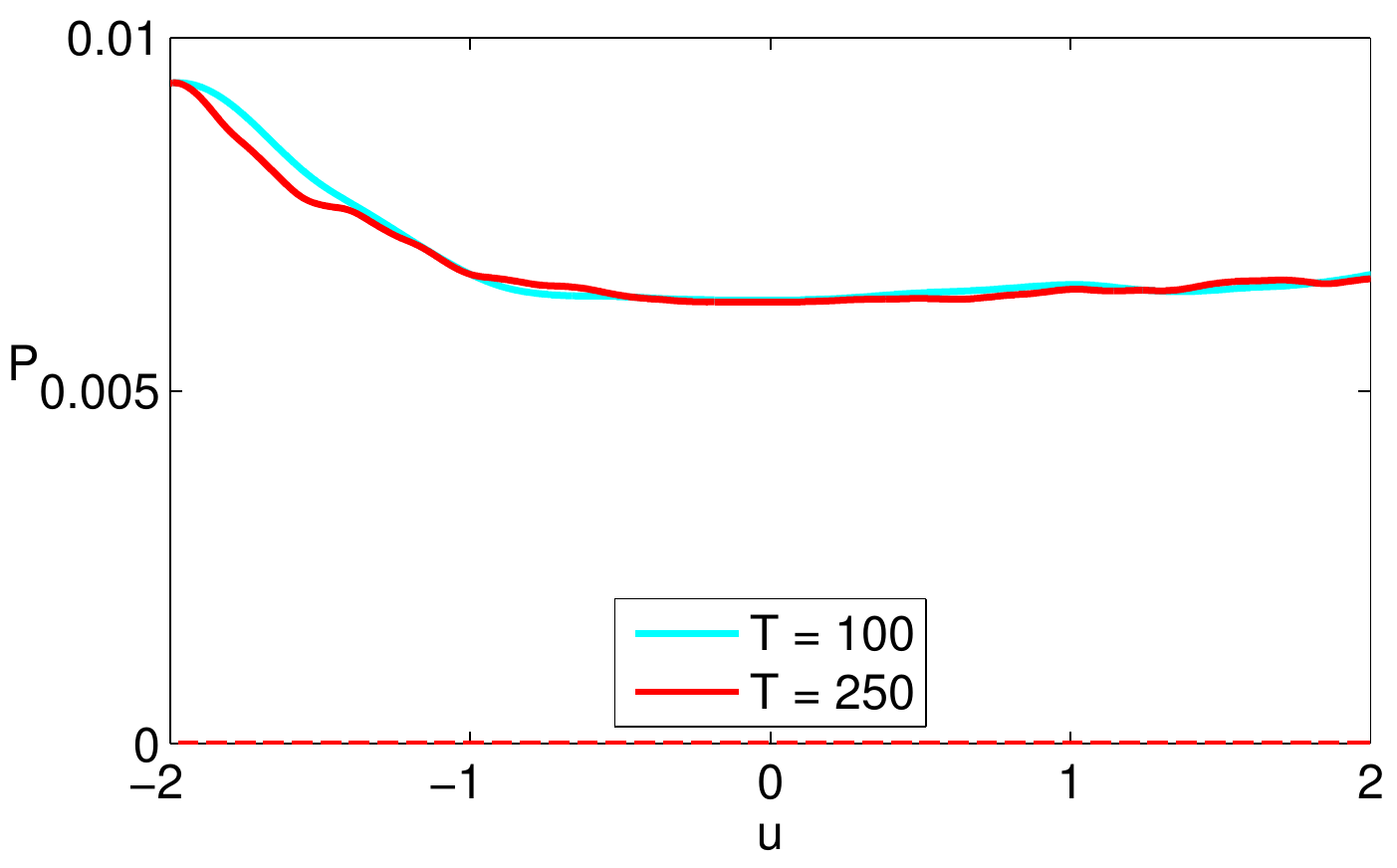}
\end{center}
\caption{Evolution of the pairing per particle during a linear ramp. The solid lines indicate the evolution of the pairing under a linear ramp from $u=-2$ to $u=2$. We see that the pairing survives even on the repulsive side. Conversely, when we start in the unpaired regime at $u=2$ and ramp to negative $u$ the pairing remains zero (dashed lines).    \label{fig:Crossover_Pairing}}
\end{figure}

We can gain an understanding of this phenomena by looking at Fig.~\ref{fig:Pairing_trap_T0}: We see that for $u=0$ the ground state undergoes a phase transition from a paired to an unpaired phase. However, as we have depicted in Fig.~\ref{fig:Crossover_Pairing}, the dynamical evolution of the system does not show this transition: When we start from $u=-2$, where we are in a paired phase, the system remains paired even when we ramp to the repulsive side (solid lines). On the other hand, the unpaired system from which we start for $u=2$ remains unpaired even when we ramp to positive values of $u$ (dashed lines).

\subsection{Dynamical process for the Hubbard model in a trap}
Finally, we investigate dynamical properties of the Hubbard model in a trap. Here, we want to investigate how the attractive Hubbard model reacts to a change of the trapping potential $V_t$. To this end, we start with the ground state of the system in a shallow trap of strength $V_t = 0.1$ and then squeeze the trap, by changing the trapping potential linearly in time, to a final trapping potential of $V_t = 0.25$. As an example, we have considered an interaction strength of $u=-6$ and we have set the chemical potential to half-filling (here: $\mu = 3$) (I). When we have reached the final value $V_t = 0.25$ we reverse the process and now decrease the trapping potential linearly to its initial value of $V_t = 0.1$ (II). 

\begin{figure}[t]
\begin{center}
\includegraphics[width = 0.95\columnwidth]{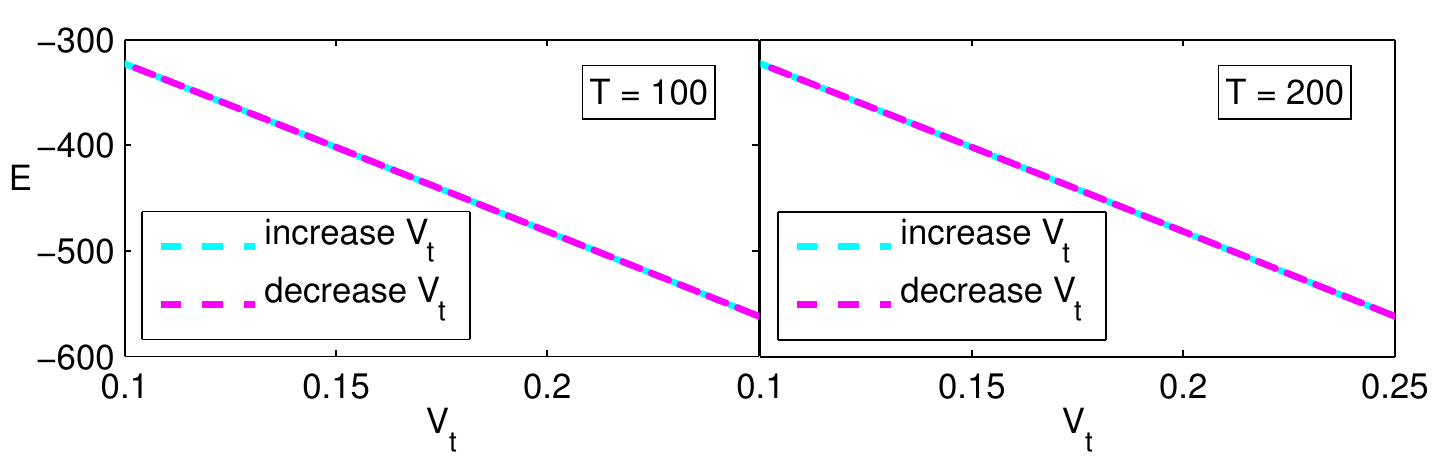}
\end{center}
\caption{Evolution of the energy for the half-filled Hubbard model with $u=-6$, $\mu =3$ (half-filling) under a change of the external trapping potential for ramping times $T_f=100$ (left) and $T_f=200$ (right).  Starting from the ground state for $V_t=0.1$ we change the trapping potential $V_t$ linearly to its final value $V_t = 0.25$ (blue line). Then we reverse the process, decreasing $V_t$ again to $V_t = 0.1$ (purple line). We see that the process is reversible.  \label{fig:Dynamic_trap_energy}}
\end{figure}

As the first question, we study if our algorithm allows for an adiabatic evolution, i.e. the energy of the system for process (I) and (II) is the same. To this end we have investigated ramping times of $T_f=100$ and $T_f = 200$, and find a reversible process in both cases (see Fig.~\ref{fig:Dynamic_trap_energy}).

\begin{figure}[t]
\begin{center}
\includegraphics[width = 0.95\columnwidth]{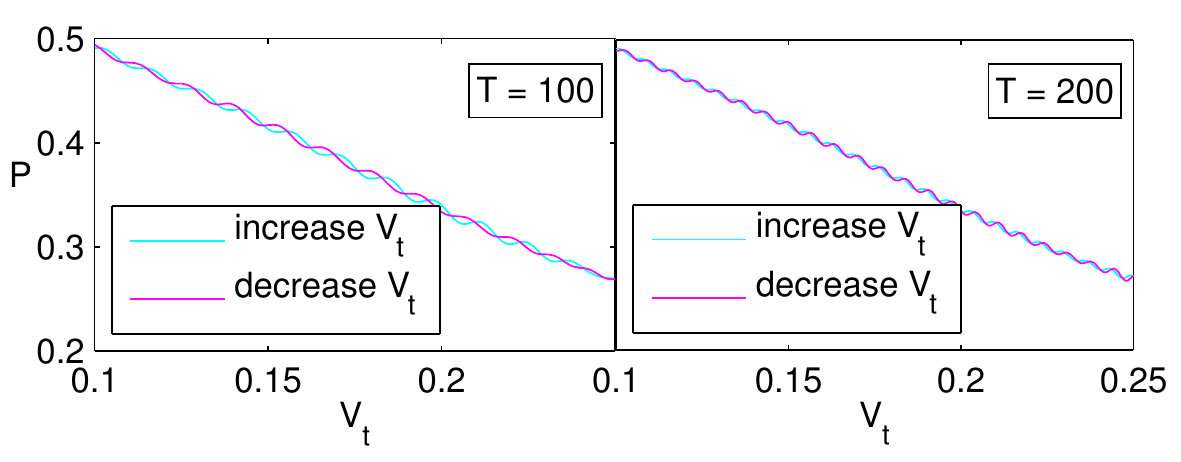}
\end{center}
\caption {Evolution of the pairing for the half-filled Hubbard model with $u=-6$, $\mu =3$ (i.e. half-filling) under a change of the external trapping potential for ramping times $T_f=100$ (left) and $T_f=200$ (right).  Starting from the ground state for $V_t=0.1$ we change the trapping potential $V_t$ linearly to its final value $V_t = 0.25$ (blue line). Then we reverse the process, decreasing $V_t$ again to $V_t = 0.1$ (purple line). We see that the process gets reversible when we go to longer ramping times, and that the evolution of the pairing gets smoother.   \label{fig:Dynamic_trap_pairing}}
\end{figure}

Next, we are interested in the evolution of the pairing under the change of the trapping potential. The result is depicted in Fig.~\ref{fig:Dynamic_trap_pairing}. We find that the process gets reversible when we go to longer ramping times, and that the evolution of the pairing gets smoother. Further, we can understand the decrease (increase) of the pairing strength with an increasing (decreasing) trapping potential by looking at Fig.~\ref{fig:Phase_diagram_T0}: The change of the trapping potential can be understood as a change of the chemical potential $\mu$ for fixed value of $u$. 

\section{Conclusion}
In this work we have formulated the gHFT for fermionic lattice systems with a two-body interaction in terms of the covariance matrix, and have derived numerical methods to find the time-evolution, the ground and thermal states in this framework. We have benchmarked our methods with the two-dimensional translationally invariant Hubbard model, and have found that our numerical methods work very well to solve the general Hartree-Fock theory. Our algorithms run fast, stable and can be applied to inhomogeneous systems, as well as to large system sizes, since all algorithms are formulated in terms of the CM, leading to a quadratic scaling in the system size. However, we also found certain limitations of our approach that are due to the fact that gHFT is intrinsically not capable of describing certain strongly correlated phases, like the Mott phase at finite temperature. 

In this respect, we believe that the numerical methods developed in this work represent a powerful tool to simulate fermionic systems in optical lattices, at least in the weakly interacting regime. Further, we have demonstrated that our algorithms can be easily applied to inhomogeneous systems, as they are realized, e.g., by the Hubbard model with an external trapping potential. 

Concerning the question if properties of the system in the thermodynamic limit can be obtained via finite-size scaling the results presented in Fig.~\ref{fig:Compare_GS_energy_QMC} and table~\ref{Crit_P} suggest that already a moderate system size of about 100 sites is enough to get a good estimate of the properties of the attractive Hubbard model in the thermodynamic limit. In addition, our numerics done for a lattice size up to $16\times 16$ show that energies can be obtained with only a slight increase in the running time, while correlation functions need some more effort, but can still be obtained in a reasonable amount of time 

Thus, in a future work, we aim at applying our approach to some particular experimental setup, which will require an extension of our current algorithms to much larger system sizes, and constitute a further benchmark of our methods.

\emph{Acknowledgments.---}This work has been supported by the EU projects QUEVADIS and COMPAS,  and the Elite Network of Bavaria program QCCC. We thank Frank Verstraete for pointing out Ref.~\cite{bach-1993} to us, and we would like to thank him, Mari Carmen Ba\~nuls  and Norbert Schuch for useful discussions.

\begin{appendix}
\section{Minimization of the energy}\label{app:Minimize_energy_Lagrange}
In order to obtain the generalized Hartree-Fock ground state we
have to solve the minimization problem
\begin{multline*}
\min_{\rho \;\mathrm{Gaussian}}E(\rho) = \min_{\rho
\;\mathrm{Gaussian}} \tr[H\rho] = \\
\min_{i\Gamma \leq \mathds{1}
}\left\{\sum_{ij} T_{ij}\Gamma_{ij} -
3\sum_{i,j,k,l}U_{ijkl}\Gamma_{ij}\Gamma_{kl}\right\}.
\end{multline*}
The condition $i\Gamma \leq \mathds{1} $ is fulfilled iff
$\mathds{1} - i\Gamma =(A+iB)(A+iB)^{\dagger}$, where $A$ and $B$
are real matrices~\cite{H&J}, from which we derive
\begin{eqnarray}
\Gamma &=& AB^T - BA^T, \label{Gamma-AB} \\
\mathds{1} &=& AA^T+BB^T. \label{1-AB}
\end{eqnarray}
Then, we can reformulate the problem as an optimization problem using Lagrange multipliers. The Lagrangian
is given by
\begin{multline}
\mathcal{L}_E= \sum_{ij}T_{ij} (AB^T -BA^T)_{ij} -\\
3\sum_{ijkl}U_{ijkl}(AB^T-BA^T)_{ij}(AB^T-BA^T)_{kl}\\
+ \sum_{i,k,l}\lambda_{kl}(A_{ki}A_{li} + B_{ki}B_{li} - \delta_{kl}),\label{eq:Lagrangian_E}
\end{multline}
where $\lambda^T = \lambda$ is the matrix of the Lagrange multipliers. The
necessary conditions for a local minimum are given by
\begin{eqnarray}
\frac{\partial \mathcal{L}_E}{\partial A_{\alpha\beta}} &=& 2 (\bar{h} B + \lambda A)_{\alpha \beta} = 0, \label{eq:Lagrange_energy_1}\\
\frac{\partial \mathcal{L}_E}{\partial B_{\alpha\beta}} &=& 2
(-\bar{h} A + \lambda B)_{\alpha \beta} = 0.\label{eq:Lagrange_energy_2}
\end{eqnarray}
From this we can obtain equations for the CM if we use Eqs. \eqref{Gamma-AB} and \eqref{1-AB}:
\begin{eqnarray}
-\bar h \Gamma + \lambda \mathds{1} &=& 0,\\
\bar h + \lambda \Gamma &=& 0.
\end{eqnarray}
Since $\Gamma^2 = -\mathds{1}$ and $\lambda^T = \lambda$ the condition $[\bar h, \Gamma] = 0$ follows immediately.

\section{Imaginary time evolution of the CM}\label{app:Proof_imaginary_time}
In this Section we give some details that lead to Eq.~\eqref{eq:Gammadot_imag}. To this end, we will show that either an evolution with the quadratic, state-dependent Hamiltonian $H_Q$ or the evolution with the interaction Hamiltonian $H(T,U)$ followed by an application of Wick's theorem lead to the desired equation.

\subsection{Imaginary time evolution with the quadratic, state-dependent Hamiltonian}
In the following, we will show how the CM evolves under an evolution in imaginary time with the quadratic, state-dependent Hamiltonian $H_Q = i\sum_{ij}\bar h_{ij}c_ic_j$, i.e., we have to calculate (c.f. Eq.~\eqref{eq:Gammadot_imag_equ})
$$ \dot \Gamma_{kl} = -i \tr\left[\{H_Q, c_kc_l\}\right] + 2 \tr[H_Q\rho]\Gamma_{kl}.$$
First, note that $\tr[H\rho] = - \tr[\Gamma \bar h]$. Next, we want to calculate $ -i \tr\left[\{H_Q, c_kc_l\}\right]$ using Wick's theorem:
\begin{multline}\label{eq:app:AC_HQ}
-i\tr[\rho(t)\{H_Q,c_kc_l\}]=\sum_{i,j}\bar{h}_{ij} \langle c_ic_jc_kc_l + c_kc_lc_ic_j\rangle \\
=2 \sum_{i,j \neq k,l} \bar h_{ij}(-\Gamma_{ij}\Gamma_{kl} + 2\Gamma_{ik}\Gamma_{jl} ) - 4 \bar h_{kl}\\
= 2 \tr[\bar{h}\Gamma]\Gamma_{kl} - 4 (\Gamma \bar{h}\Gamma + \bar{h})_{kl},
\end{multline}
and we obtain $ \dot \Gamma= -4(\Gamma \bar h \Gamma + \bar h)$. 
%
\subsection{Evolution with the interaction Hamiltonian}
In the following, we will show how the CM evolves under an evolution in imaginary time with the interaction Hamiltonian $H(T,U) = i\sum_{ij}T_{ij}c_ic_j + \sum_{ijkl}U_{ijkl}c_ic_jc_kc_l$, i.e. we have to evaluate (c.f. Eq.~\eqref{eq:Gammadot_imag_equ})
$$ \dot \Gamma_{kl} = -i \tr\left[\{H(T,U), c_kc_l\}\right] + 2 \tr[H(T,U)\rho]\Gamma_{kl}.$$
We split the Hamiltonian $H(T,U)$ in two terms,  $H_q = i\sum_{kl} T_{kl}c_kc_l$ and $H_I = \sum_{klmn}U_{klmn}c_kc_lc_mc_n$, and perform the calculation for both terms independently. First, the calculation for $H_q$ is the same as for $H_Q$, so that we can deduce from Eq.~\eqref{eq:app:AC_HQ} 
\begin{equation}
-i\tr[\rho(t)\{H_q,c_kc_l\}] = 2 \tr[T\Gamma]\Gamma_{kl} - 4 (\Gamma T \Gamma +T)_{kl}.
\end{equation}
Next, we calculate the contribution of the interaction term. First, 
\begin{multline*}
-i \tr[\{H_I,c_kc_l\}\rho(t)] =\\
 -2 i \sum_{i,j,m, n \neq k, l}U_{ijmn}  \langle c_ic_jc_mc_nc_kc_l\rangle - 24 \tr_B[U\Gamma]_{kl}. 
\end{multline*}
Next, using Wick's theorem, we find 
\begin{multline*}
-2 i \sum_{i,j,m, n \neq k, l}U_{ijmn}  \langle c_ic_jc_mc_nc_kc_l\rangle = \\
 \sum_{i,j,m, n \neq k, l}U_{ijmn} \left(6\Gamma_{ij}\Gamma_{mn}\Gamma_{kl} - 24 \Gamma_{ik}\Gamma_{jm}\Gamma_{nl}\right).
\end{multline*}
We treat the two terms independently, starting with $\sum_{i,j,m, n \neq k, l}U_{ijmn} \Gamma_{ij}\Gamma_{mn}$:
\begin{multline*}
\sum_{i,j,m, n \neq k, l}U_{ijmn} \Gamma_{ij}\Gamma_{mn} = \\
\tr[\Gamma \tr_B[U\Gamma]] - 4 (\tr_B[U\Gamma]\Gamma)_{kk}  - 4 (\tr_B[U\Gamma]\Gamma)_{ll}\\
- 8 \sum_{i,j}U_{ijkl}\Gamma_{ik}\Gamma_{jl} - 4 \tr_B[U\Gamma]_{kl}\Gamma_{kl}.
\end{multline*}
Next, we calculate $\sum_{i,j,m, n \neq k, l}U_{ijmn} \Gamma_{ik}\Gamma_{jm}\Gamma_{nl}$:
\begin{multline*}
\sum_{i,j,m, n \neq k, l}U_{ijmn} \Gamma_{ik}\Gamma_{jm}\Gamma_{nl} = \\
(\Gamma \tr_B[U\Gamma]\Gamma)_{kl} - (\tr_B[U\Gamma]\Gamma)_{kk} \Gamma_{kl} - (\tr_B[U\Gamma]\Gamma)_{ll} \Gamma_{kl}\\
- \tr_B[U\Gamma]_{kl}\Gamma_{kl}^2 - 2 \sum_{i,n}U_{klin}\Gamma_{ik}\Gamma_{nl}\Gamma_{kl}
\end{multline*}
Thus, we obtain 
\begin{multline*}
-i \tr[\rho\{H_I,c_kc_l\}] = \\
-24 \tr[U\Gamma]_{kl} - 24 (\Gamma\tr[U\Gamma]\Gamma)_{kl} + 6
\tr[\Gamma\tr[U\Gamma]]\Gamma_{kl}.
\end{multline*}
Hence, $-i \tr[\rho\{H,c_kc_l\}] = -2\tr[H\rho]\Gamma_{kl} - 4(\Gamma \bar h \Gamma + \bar h)$, which leads to Eq. \eqref{eq:Gammadot_imag}.
%

\section{Minimization of the free energy}\label{app:Minimize_Free_Energy}
We want to minimize
$$
\min_{\rho \;\mathrm{Gaussian}} F(\rho) = \min_{\rho \;\mathrm{Gaussian}} \left\{E(\rho) -
\beta^{-1}S(\rho)\right\},
$$
with the help of Lagrange multipliers. An expression for the energy has already been given in Eq.~\eqref{eq:minE}. The entropy $S = - \tr[\rho
\ln \rho]$ is given by $ S = M \ln 2 -
\frac{1}{2}\tr\left[(\mathds{1}+i\Gamma)\ln(\mathds{1}+i\Gamma)\right]$ \cite{wolf:010404}, so that we have to solve the following optimization problem:
\begin{multline*}
\min_{\rho \;\mathrm{Gaussian}}F(\rho)=\\
\min_{i\Gamma \leq \mathds{1}}\bigg\{ -\tr[(T + 3 \tr_B[U\Gamma])
\Gamma]\\
- \beta^{-1}\left(M\ln 2
-\frac{1}{2}\tr[(\mathds{1}+i\Gamma)\ln(\mathds{1}+i\Gamma)]\right)\bigg\}.
\end{multline*}
As we have already stated in Appendix \ref{app:Minimize_energy_Lagrange}, the condition $i\Gamma \leq \mathds{1} $ is fulfilled iff
$\mathds{1} - i\Gamma =(A+iB)(A+iB)^{\dagger}$, where $A$ and $B$
are real matrices~\cite{H&J}, from which we derive
\begin{eqnarray}
\Gamma &=& AB^T - BA^T, \label{Gamma-AB_c} \\
\mathds{1} &=& AA^T+BB^T. \label{1-AB_c}
\end{eqnarray}
Thus, we can reformulate the minimization of the free energy as an optimization problem with Lagrange multipliers. The Lagrangian is of the form
\begin{align}
\mathcal{L}_S &=& F(A,B) + \sum_{i,k,l}\lambda_{kl}(A_{ki}A_{li} + B_{ki}B_{li} - \delta_{kl}),\label{eq:Lagrangian_F}
\end{align}
where $\lambda = \lambda^T$ is the matrix of Lagrange multipliers. Now, $\frac{\partial \mathcal L_S}{\partial A_{mn}} =\frac{\partial \mathcal L_S}{\partial \Gamma_{kl}}\frac{\partial  \Gamma_{kl}}{\partial A_{mn}}$  (resp. $B$), and we calculate first $\frac{\partial F}{\partial \Gamma_{kl}}$. To this end, we consider the entropy term, $\tr[(\mathds{1}+i\Gamma)\ln(\mathds{1}+i\Gamma)]$: We us  $\ln
(\mathds{1} - X) = -\sum_{k=1}^{\infty}\frac{1}{k} X^k$ (see e.g.~\cite{H&J2}) leading to
$$\tr[(\mathds{1}+i\Gamma)\ln (\mathds{1}+i\Gamma)] = - \sum_{k,l} \frac{1}{k}(-i)^k (\Gamma^k + i
\Gamma^{k+1})_{ll}.$$
Then
\begin{multline}
\frac{\partial }{\partial \Gamma_{kl}} \tr[(\mathds{1}+i\Gamma)\ln(\mathds{1}+i\Gamma)]
=i\left[\mathds{1} + \ln (\mathds{1} - i\Gamma)\right]_{kl},
\end{multline}
where we have made use of the formula $(\mathds{1} - X)^{-1} = \sum_{k=0}^{\infty}X^k$ ~\cite{H&J2}. Since $[\ln (\mathds{1} - i\Gamma)]^T = \ln (\mathds{1} + i\Gamma)$, we arrive at
\begin{align*}
\frac{\partial \mathcal L_S}{\partial A_{mn}} &= 2(h_F B + \lambda A) = 0,\\
\frac{\partial \mathcal L_S}{\partial B_{mn}} &= 2(-h_F A + \lambda B) = 0,\\
h_F(\Gamma) &= \bar{h} - \frac{i}{4\beta}\ln \frac{\mathds{1} + i
\Gamma}{\mathds{1} - i \Gamma}.
\end{align*}
Then, using Eqs.~\eqref{Gamma-AB_c} and \eqref{1-AB_c} we obtain the following necessary conditions for a minimal entropy:
\begin{eqnarray}
[h_F(\Gamma),\Gamma] &=& 0, \label{HFcom}\\
h_F(\Gamma) (\mathds{1} + \Gamma^2) &=& 0. \label{HF0}
\end{eqnarray}
\end{appendix}


\begin{thebibliography}{44}
\expandafter\ifx\csname natexlab\endcsname\relax\def\natexlab#1{#1}\fi
\expandafter\ifx\csname bibnamefont\endcsname\relax
  \def\bibnamefont#1{#1}\fi
\expandafter\ifx\csname bibfnamefont\endcsname\relax
  \def\bibfnamefont#1{#1}\fi
\expandafter\ifx\csname citenamefont\endcsname\relax
  \def\citenamefont#1{#1}\fi
\expandafter\ifx\csname url\endcsname\relax
  \def\url#1{\texttt{#1}}\fi
\expandafter\ifx\csname urlprefix\endcsname\relax\def\urlprefix{URL }\fi
\providecommand{\bibinfo}[2]{#2}
\providecommand{\eprint}[2][]{\url{#2}}

\bibitem[{\citenamefont{Dalfovo et~al.}(1999)\citenamefont{Dalfovo, Giorgini,
  Pitaevskii, and Stringari}}]{Stringari_Bosons}
\bibinfo{author}{\bibfnamefont{F.}~\bibnamefont{Dalfovo}},
  \bibinfo{author}{\bibfnamefont{S.}~\bibnamefont{Giorgini}},
  \bibinfo{author}{\bibfnamefont{L.~P.} \bibnamefont{Pitaevskii}},
  \bibnamefont{and}
  \bibinfo{author}{\bibfnamefont{S.}~\bibnamefont{Stringari}},
  \bibinfo{journal}{Rev. Mod. Phys.} \textbf{\bibinfo{volume}{71}},
  \bibinfo{pages}{463} (\bibinfo{year}{1999}).

\bibitem[{\citenamefont{Giorgini et~al.}(2008)\citenamefont{Giorgini,
  Pitaevskii, and Stringari}}]{Stringari_Fermions}
\bibinfo{author}{\bibfnamefont{S.}~\bibnamefont{Giorgini}},
  \bibinfo{author}{\bibfnamefont{L.~P.} \bibnamefont{Pitaevskii}},
  \bibnamefont{and}
  \bibinfo{author}{\bibfnamefont{S.}~\bibnamefont{Stringari}},
  \bibinfo{journal}{Rev. Mod. Phys.} \textbf{\bibinfo{volume}{80}},
  \bibinfo{pages}{1215} (\bibinfo{year}{2008}).

\bibitem[{\citenamefont{Bloch et~al.}(2008)\citenamefont{Bloch, Dalibard, and
  Zwerger}}]{bloch-2008-80}
\bibinfo{author}{\bibfnamefont{I.}~\bibnamefont{Bloch}},
  \bibinfo{author}{\bibfnamefont{J.}~\bibnamefont{Dalibard}}, \bibnamefont{and}
  \bibinfo{author}{\bibfnamefont{W.}~\bibnamefont{Zwerger}},
  \bibinfo{journal}{Reviews of Modern Physics} \textbf{\bibinfo{volume}{80}},
  \bibinfo{pages}{885} (\bibinfo{year}{2008}).

\bibitem[{\citenamefont{Pitaevskii and Stringari}(2003)}]{Piataevskii}
\bibinfo{author}{\bibfnamefont{L.}~\bibnamefont{Pitaevskii}} \bibnamefont{and}
  \bibinfo{author}{\bibfnamefont{S.}~\bibnamefont{Stringari}},
  \emph{\bibinfo{title}{Bose Einstein Condensation}}
  (\bibinfo{publisher}{Clarendon Press, Oxford}, \bibinfo{year}{2003}).

\bibitem[{\citenamefont{Fetter and Walecka}(1980)}]{fetter:walecka}
\bibinfo{author}{\bibfnamefont{A.~L.} \bibnamefont{Fetter}} \bibnamefont{and}
  \bibinfo{author}{\bibfnamefont{J.~D.} \bibnamefont{Walecka}},
  \emph{\bibinfo{title}{Theoretical Mechanics of Particles and Continua}}
  (\bibinfo{year}{1980}).

\bibitem[{\citenamefont{Barankov et~al.}(2004)\citenamefont{Barankov, Levitov,
  and Spivak}}]{barankov-2004-93}
\bibinfo{author}{\bibfnamefont{R.~A.} \bibnamefont{Barankov}},
  \bibinfo{author}{\bibfnamefont{L.~S.} \bibnamefont{Levitov}},
  \bibnamefont{and} \bibinfo{author}{\bibfnamefont{B.~Z.}
  \bibnamefont{Spivak}}, \bibinfo{journal}{Physical Review Letters}
  \textbf{\bibinfo{volume}{93}}, \bibinfo{pages}{160401}
  (\bibinfo{year}{2004}).

\bibitem[{\citenamefont{Barankov and
  Levitov}(2006{\natexlab{a}})}]{barankov-2006-73}
\bibinfo{author}{\bibfnamefont{R.~A.} \bibnamefont{Barankov}} \bibnamefont{and}
  \bibinfo{author}{\bibfnamefont{L.~S.} \bibnamefont{Levitov}},
  \bibinfo{journal}{Physical Review A} \textbf{\bibinfo{volume}{73}},
  \bibinfo{pages}{033614} (\bibinfo{year}{2006}{\natexlab{a}}).

\bibitem[{\citenamefont{Barankov and
  Levitov}(2006{\natexlab{b}})}]{Barankov-PhysRevA.73.033614}
\bibinfo{author}{\bibfnamefont{R.~A.} \bibnamefont{Barankov}} \bibnamefont{and}
  \bibinfo{author}{\bibfnamefont{L.~S.} \bibnamefont{Levitov}},
  \bibinfo{journal}{Phys. Rev. A} \textbf{\bibinfo{volume}{73}},
  \bibinfo{pages}{033614} (\bibinfo{year}{2006}{\natexlab{b}}).

\bibitem[{\citenamefont{Hastings and Levitov}(2008)}]{hastings-2008-BCS}
\bibinfo{author}{\bibfnamefont{M.~B.} \bibnamefont{Hastings}} \bibnamefont{and}
  \bibinfo{author}{\bibfnamefont{L.~S.} \bibnamefont{Levitov}},
  \emph{\bibinfo{title}{Synchronization and dephasing of many-body states in
  optical lattices}} (\bibinfo{year}{2008}).

\bibitem[{\citenamefont{Massel and Penna}(2006)}]{Massel}
\bibinfo{author}{\bibfnamefont{F.~P.} \bibnamefont{Massel}} \bibnamefont{and}
  \bibinfo{author}{\bibfnamefont{V.}~\bibnamefont{Penna}},
  \bibinfo{journal}{Journal of Physics B: Atomic, Molecular and Optical
  Physics} \textbf{\bibinfo{volume}{39}}, \bibinfo{pages}{S143}
  (\bibinfo{year}{2006}).

\bibitem[{\citenamefont{Bach et~al.}(1994)\citenamefont{Bach, Lieb, and
  Solovej}}]{bach-1993}
\bibinfo{author}{\bibfnamefont{V.}~\bibnamefont{Bach}},
  \bibinfo{author}{\bibfnamefont{E.~H.} \bibnamefont{Lieb}}, \bibnamefont{and}
  \bibinfo{author}{\bibfnamefont{J.~P.} \bibnamefont{Solovej}},
  \bibinfo{journal}{J. Stat. Phys.} \textbf{\bibinfo{volume}{76}},
  \bibinfo{pages}{3} (\bibinfo{year}{1994}).

\bibitem[{\citenamefont{Bravyi}(2005)}]{LinearOptics}
\bibinfo{author}{\bibfnamefont{S.}~\bibnamefont{Bravyi}},
  \bibinfo{journal}{Quant. Inf. Comput.} \textbf{\bibinfo{volume}{5}},
  \bibinfo{pages}{216} (\bibinfo{year}{2005}).

\bibitem[{\citenamefont{Varney et~al.}(2009)\citenamefont{Varney, Lee, Bai,
  Chiesa, Jarrell, and Scalettar}}]{Scarletter}
\bibinfo{author}{\bibfnamefont{C.~N.} \bibnamefont{Varney}},
  \bibinfo{author}{\bibfnamefont{C.-R.} \bibnamefont{Lee}},
  \bibinfo{author}{\bibfnamefont{Z.~J.} \bibnamefont{Bai}},
  \bibinfo{author}{\bibfnamefont{S.}~\bibnamefont{Chiesa}},
  \bibinfo{author}{\bibfnamefont{M.}~\bibnamefont{Jarrell}}, \bibnamefont{and}
  \bibinfo{author}{\bibfnamefont{R.~T.} \bibnamefont{Scalettar}},
  \bibinfo{journal}{Phys. Rev. B} \textbf{\bibinfo{volume}{80}},
  \bibinfo{pages}{075116} (\bibinfo{year}{2009}).

\bibitem[{\citenamefont{Paiva et~al.}(2009)\citenamefont{Paiva, Scalettar,
  Randeria, and Trivedi}}]{trivedi-2009}
\bibinfo{author}{\bibfnamefont{T.}~\bibnamefont{Paiva}},
  \bibinfo{author}{\bibfnamefont{R.}~\bibnamefont{Scalettar}},
  \bibinfo{author}{\bibfnamefont{M.}~\bibnamefont{Randeria}}, \bibnamefont{and}
  \bibinfo{author}{\bibfnamefont{N.}~\bibnamefont{Trivedi}},
  \eprint{arXiv.org:0906.2141}
  (\bibinfo{year}{2009}).

\bibitem[{\citenamefont{Corney and Drummond}(2004)}]{corney-2004-93}
\bibinfo{author}{\bibfnamefont{J.~F.} \bibnamefont{Corney}} \bibnamefont{and}
  \bibinfo{author}{\bibfnamefont{P.~D.} \bibnamefont{Drummond}},
  \bibinfo{journal}{Physical Review Letters} \textbf{\bibinfo{volume}{93}},
  \bibinfo{pages}{260401} (\bibinfo{year}{2004}).

\bibitem[{\citenamefont{Kraus et~al.}(2009)\citenamefont{Kraus, Schuch,
  Verstraete, and Cirac}}]{kraus-2009}
\bibinfo{author}{\bibfnamefont{C.~V.} \bibnamefont{Kraus}},
  \bibinfo{author}{\bibfnamefont{N.}~\bibnamefont{Schuch}},
  \bibinfo{author}{\bibfnamefont{F.}~\bibnamefont{Verstraete}},
  \bibnamefont{and} \bibinfo{author}{\bibfnamefont{J.~I.} \bibnamefont{Cirac}}, \bibinfo{journal}{Physical    Review A} \textbf{\bibinfo{volume}{81}},
  \bibinfo{pages}{052338} (\bibinfo{year}{2010}).
  
 
\bibitem[{\citenamefont{Corboz et~al.}(2009)\citenamefont{Corboz, Orus, Bauer,
  and Vidal}}]{corboz-2009}
\bibinfo{author}{\bibfnamefont{P.}~\bibnamefont{Corboz}},
  \bibinfo{author}{\bibfnamefont{R.}~\bibnamefont{Orus}},
  \bibinfo{author}{\bibfnamefont{B.}~\bibnamefont{Bauer}}, \bibnamefont{and}
  \bibinfo{author}{\bibfnamefont{G.}~\bibnamefont{Vidal}},
  \bibinfo{journal}{Physical Review B} \textbf{\bibinfo{volume}{81}},
  \bibinfo{pages}{165104 } (\bibinfo{year}{2010}).

\bibitem[{\citenamefont{Pineda et~al.}(2009)\citenamefont{Pineda, Barthel, and
  Eisert}}]{pineda-2009}
\bibinfo{author}{\bibfnamefont{C.}~\bibnamefont{Pineda}},
  \bibinfo{author}{\bibfnamefont{T.}~\bibnamefont{Barthel}}, \bibnamefont{and}
  \bibinfo{author}{\bibfnamefont{J.}~\bibnamefont{Eisert}},
  \eprint{arXiv.org:0905.0669} (\bibinfo{year}{2009}).

\bibitem[{\citenamefont{Corboz and Vidal}(2009)}]{corboz-2009-80}
\bibinfo{author}{\bibfnamefont{P.}~\bibnamefont{Corboz}} \bibnamefont{and}
  \bibinfo{author}{\bibfnamefont{G.}~\bibnamefont{Vidal}},
  \bibinfo{journal}{Physical Review B} \textbf{\bibinfo{volume}{80}},
  \bibinfo{pages}{165129} (\bibinfo{year}{2009}).

\bibitem[{\citenamefont{Barthel et~al.}(2009)\citenamefont{Barthel, Pineda, and
  Eisert}}]{barthel-2009-80}
\bibinfo{author}{\bibfnamefont{T.}~\bibnamefont{Barthel}},
  \bibinfo{author}{\bibfnamefont{C.}~\bibnamefont{Pineda}}, \bibnamefont{and}
  \bibinfo{author}{\bibfnamefont{J.}~\bibnamefont{Eisert}},
  \bibinfo{journal}{Physical Review A} \textbf{\bibinfo{volume}{80}},
  \bibinfo{pages}{042333} (\bibinfo{year}{2009}).

\bibitem[{\citenamefont{Li et~al.}(20109)\citenamefont{Li, Shi, and
  Zhou}}]{Zhou-2010}
\bibinfo{author}{\bibfnamefont{S.}~\bibnamefont{Li}},
  \bibinfo{author}{\bibfnamefont{Q.}~\bibnamefont{Shi}}, \bibnamefont{and}
  \bibinfo{author}{\bibfnamefont{H.}~\bibnamefont{Zhou}},
  \eprint{arXiv.org:1001.3343}(\bibinfo{year}{2010}).

\bibitem[{\citenamefont{Hofstetter et~al.}(2002)\citenamefont{Hofstetter,
  Cirac, Zoller, Demler, and Lukin}}]{Hofstetter}
\bibinfo{author}{\bibfnamefont{W.}~\bibnamefont{Hofstetter}},
  \bibinfo{author}{\bibfnamefont{J.~I.} \bibnamefont{Cirac}},
  \bibinfo{author}{\bibfnamefont{P.}~\bibnamefont{Zoller}},
  \bibinfo{author}{\bibfnamefont{E.}~\bibnamefont{Demler}}, \bibnamefont{and}
  \bibinfo{author}{\bibfnamefont{M.~D.} \bibnamefont{Lukin}},
  \bibinfo{journal}{Phys. Rev. Lett.} \textbf{\bibinfo{volume}{89}},
  \bibinfo{pages}{220407} (\bibinfo{year}{2002}).

\bibitem[{\citenamefont{Hubbard}(1963)}]{Hubbard_original}
\bibinfo{author}{\bibfnamefont{J.}~\bibnamefont{Hubbard}},
  \bibinfo{journal}{Proceedings of the Royal Society of London. Series A,
  Mathematical and Physical Sciences} \textbf{\bibinfo{volume}{276}},
  \bibinfo{pages}{238} (\bibinfo{year}{1963}), ISSN \bibinfo{issn}{00804630}.

\bibitem[{\citenamefont{Anderson}(1987)}]{Anderson_highTc}
\bibinfo{author}{\bibfnamefont{P.}~\bibnamefont{Anderson}},
  \bibinfo{journal}{Science} \textbf{\bibinfo{volume}{235}},
  \bibinfo{pages}{1196} (\bibinfo{year}{1987}).

\bibitem[{\citenamefont{Bloch and Messiah}(1962)}]{BlochMessiah}
\bibinfo{author}{\bibfnamefont{C.}~\bibnamefont{Bloch}} \bibnamefont{and}
  \bibinfo{author}{\bibfnamefont{A.}~\bibnamefont{Messiah}},
  \bibinfo{journal}{Nuclear Physics} \textbf{\bibinfo{volume}{39}},
  \bibinfo{pages}{95} (\bibinfo{year}{1962}).

\bibitem[{\citenamefont{Kraus et~al.}(2007)\citenamefont{Kraus, Wolf, and
  Cirac}}]{kraus:022303}
\bibinfo{author}{\bibfnamefont{C.~V.} \bibnamefont{Kraus}},
  \bibinfo{author}{\bibfnamefont{M.~M.} \bibnamefont{Wolf}}, \bibnamefont{and}
  \bibinfo{author}{\bibfnamefont{J.~I.} \bibnamefont{Cirac}},
  \bibinfo{journal}{Physical Review A} \textbf{\bibinfo{volume}{75}},
  \bibinfo{eid}{022303} (pages~\bibinfo{numpages}{9}) (\bibinfo{year}{2007}).

\bibitem[{\citenamefont{Micnas et~al.}(1990)\citenamefont{Micnas, Ranninger,
  and Robaszkiewicz}}]{Micnas_RMP}
\bibinfo{author}{\bibfnamefont{R.}~\bibnamefont{Micnas}},
  \bibinfo{author}{\bibfnamefont{J.}~\bibnamefont{Ranninger}},
  \bibnamefont{and}
  \bibinfo{author}{\bibfnamefont{S.}~\bibnamefont{Robaszkiewicz}},
  \bibinfo{journal}{Rev. Mod. Phys.} \textbf{\bibinfo{volume}{62}},
  \bibinfo{pages}{113} (\bibinfo{year}{1990}).

\bibitem[{\citenamefont{Scalapino}(1995{\natexlab{a}})}]{Scalapino_BCS}
\bibinfo{author}{\bibfnamefont{D.~J.} \bibnamefont{Scalapino}},
  \bibinfo{journal}{Physics Reports} \textbf{\bibinfo{volume}{250}},
  \bibinfo{pages}{329 } (\bibinfo{year}{1995}{\natexlab{a}}), ISSN
  \bibinfo{issn}{0370-1573}.

\bibitem[{\citenamefont{Dopf et~al.}(1992)\citenamefont{Dopf, Wagner,
  Dieterich, and Muramatsu}}]{Muramatsu_sign}
\bibinfo{author}{\bibfnamefont{G.}~\bibnamefont{Dopf}},
  \bibinfo{author}{\bibfnamefont{J.}~\bibnamefont{Wagner}},
  \bibinfo{author}{\bibfnamefont{P.}~\bibnamefont{Dieterich}},
  \bibnamefont{and}
  \bibinfo{author}{\bibfnamefont{A.}~\bibnamefont{Muramatsu}},
  \bibinfo{journal}{Helvetica Physica Acta} \textbf{\bibinfo{volume}{65}},
  \bibinfo{pages}{257} (\bibinfo{year}{1992}).

\bibitem[{\citenamefont{Giamarchi and Lhuillier}(1991)}]{Giamarchi}
\bibinfo{author}{\bibfnamefont{T.}~\bibnamefont{Giamarchi}} \bibnamefont{and}
  \bibinfo{author}{\bibfnamefont{C.}~\bibnamefont{Lhuillier}},
  \bibinfo{journal}{Phys. Rev. B} \textbf{\bibinfo{volume}{43}},
  \bibinfo{pages}{12943} (\bibinfo{year}{1991}).

\bibitem[{\citenamefont{Hirsch}(1985)}]{Hirsch}
\bibinfo{author}{\bibfnamefont{J.~E.} \bibnamefont{Hirsch}},
  \bibinfo{journal}{Phys. Rev. B} \textbf{\bibinfo{volume}{31}},
  \bibinfo{pages}{4403} (\bibinfo{year}{1985}).

\bibitem[{\citenamefont{Bulut et~al.}(1994)\citenamefont{Bulut, Scalapino, and
  White}}]{White_QMC_94}
\bibinfo{author}{\bibfnamefont{N.}~\bibnamefont{Bulut}},
  \bibinfo{author}{\bibfnamefont{D.~J.} \bibnamefont{Scalapino}},
  \bibnamefont{and} \bibinfo{author}{\bibfnamefont{S.~R.} \bibnamefont{White}},
  \bibinfo{journal}{Phys. Rev. Lett.} \textbf{\bibinfo{volume}{73}},
  \bibinfo{pages}{748} (\bibinfo{year}{1994}).

\bibitem[{\citenamefont{Maier et~al.}(2005{\natexlab{a}})\citenamefont{Maier,
  Jarrell, Schulthess, Kent, and White}}]{White_QMC_95}
\bibinfo{author}{\bibfnamefont{T.~A.} \bibnamefont{Maier}},
  \bibinfo{author}{\bibfnamefont{M.}~\bibnamefont{Jarrell}},
  \bibinfo{author}{\bibfnamefont{T.~C.} \bibnamefont{Schulthess}},
  \bibinfo{author}{\bibfnamefont{P.~R.~C.} \bibnamefont{Kent}},
  \bibnamefont{and} \bibinfo{author}{\bibfnamefont{J.~B.} \bibnamefont{White}},
  \bibinfo{journal}{Phys. Rev. Lett.} \textbf{\bibinfo{volume}{95}},
  \bibinfo{pages}{237001} (\bibinfo{year}{2005}{\natexlab{a}}).

\bibitem[{\citenamefont{Chang and Zhang}(2008)}]{Zhang_QMC}
\bibinfo{author}{\bibfnamefont{C.-C.} \bibnamefont{Chang}} \bibnamefont{and}
  \bibinfo{author}{\bibfnamefont{S.}~\bibnamefont{Zhang}},
  \bibinfo{journal}{Phys. Rev. B} \textbf{\bibinfo{volume}{78}},
  \bibinfo{pages}{165101} (\bibinfo{year}{2008}).

\bibitem[{\citenamefont{Haas et~al.}(1995)\citenamefont{Haas, Moreo, and
  Dagotto}}]{Dagotto_QMC}
\bibinfo{author}{\bibfnamefont{S.}~\bibnamefont{Haas}},
  \bibinfo{author}{\bibfnamefont{A.}~\bibnamefont{Moreo}}, \bibnamefont{and}
  \bibinfo{author}{\bibfnamefont{E.}~\bibnamefont{Dagotto}},
  \bibinfo{journal}{Phys. Rev. Lett.} \textbf{\bibinfo{volume}{74}},
  \bibinfo{pages}{4281} (\bibinfo{year}{1995}).

\bibitem[{\citenamefont{Georges et~al.}(1996)\citenamefont{Georges, Kotliar,
  Krauth, and Rozenberg}}]{Rozenberg_QMC}
\bibinfo{author}{\bibfnamefont{A.}~\bibnamefont{Georges}},
  \bibinfo{author}{\bibfnamefont{G.}~\bibnamefont{Kotliar}},
  \bibinfo{author}{\bibfnamefont{W.}~\bibnamefont{Krauth}}, \bibnamefont{and}
  \bibinfo{author}{\bibfnamefont{M.~J.} \bibnamefont{Rozenberg}},
  \bibinfo{journal}{Rev. Mod. Phys.} \textbf{\bibinfo{volume}{68}},
  \bibinfo{pages}{13} (\bibinfo{year}{1996}).

\bibitem[{\citenamefont{Preuss et~al.}(1997)\citenamefont{Preuss, Hanke,
  Gr\"ober, and Evertz}}]{Evertz_QMC}
\bibinfo{author}{\bibfnamefont{R.}~\bibnamefont{Preuss}},
  \bibinfo{author}{\bibfnamefont{W.}~\bibnamefont{Hanke}},
  \bibinfo{author}{\bibfnamefont{C.}~\bibnamefont{Gr\"ober}}, \bibnamefont{and}
  \bibinfo{author}{\bibfnamefont{H.~G.} \bibnamefont{Evertz}},
  \bibinfo{journal}{Phys. Rev. Lett.} \textbf{\bibinfo{volume}{79}},
  \bibinfo{pages}{1122} (\bibinfo{year}{1997}).

\bibitem[{\citenamefont{Gr\"ober et~al.}(2000)\citenamefont{Gr\"ober, Eder, and
  Hanke}}]{Hanke_QMC}
\bibinfo{author}{\bibfnamefont{C.}~\bibnamefont{Gr\"ober}},
  \bibinfo{author}{\bibfnamefont{R.}~\bibnamefont{Eder}}, \bibnamefont{and}
  \bibinfo{author}{\bibfnamefont{W.}~\bibnamefont{Hanke}},
  \bibinfo{journal}{Phys. Rev. B} \textbf{\bibinfo{volume}{62}},
  \bibinfo{pages}{4336} (\bibinfo{year}{2000}).

\bibitem[{\citenamefont{Maier et~al.}(2005{\natexlab{b}})\citenamefont{Maier,
  Jarrell, Pruschke, and Hettler}}]{Hettler_QMC}
\bibinfo{author}{\bibfnamefont{T.}~\bibnamefont{Maier}},
  \bibinfo{author}{\bibfnamefont{M.}~\bibnamefont{Jarrell}},
  \bibinfo{author}{\bibfnamefont{T.}~\bibnamefont{Pruschke}}, \bibnamefont{and}
  \bibinfo{author}{\bibfnamefont{M.~H.} \bibnamefont{Hettler}},
  \bibinfo{journal}{Rev. Mod. Phys.} \textbf{\bibinfo{volume}{77}},
  \bibinfo{pages}{1027} (\bibinfo{year}{2005}{\natexlab{b}}).

\bibitem[{\citenamefont{Dar\'e et~al.}(2007)\citenamefont{Dar\'e, Raymond,
  Albinet, and Tremblay}}]{Tremblay_QMC}
\bibinfo{author}{\bibfnamefont{A.-M.} \bibnamefont{Dar\'e}},
  \bibinfo{author}{\bibfnamefont{L.}~\bibnamefont{Raymond}},
  \bibinfo{author}{\bibfnamefont{G.}~\bibnamefont{Albinet}}, \bibnamefont{and}
  \bibinfo{author}{\bibfnamefont{A.-M.~S.} \bibnamefont{Tremblay}},
  \bibinfo{journal}{Phys. Rev. B} \textbf{\bibinfo{volume}{76}},
  \bibinfo{pages}{064402} (\bibinfo{year}{2007}).

\bibitem[{\citenamefont{Scalapino}(1995{\natexlab{b}})}]{ScalapinoMottGap}
\bibinfo{author}{\bibfnamefont{D.~J.} \bibnamefont{Scalapino}},
  \bibinfo{journal}{Physics Reports} \textbf{\bibinfo{volume}{250}},
  \bibinfo{pages}{329 } (\bibinfo{year}{1995}{\natexlab{b}}), ISSN
  \bibinfo{issn}{0370-1573}.

\bibitem[{\citenamefont{Horn and Johnson}(1985{\natexlab{a}})}]{H&J}
\bibinfo{author}{\bibfnamefont{R.}~\bibnamefont{Horn}} \bibnamefont{and}
  \bibinfo{author}{\bibfnamefont{C.}~\bibnamefont{Johnson}},
  \emph{\bibinfo{title}{Matrix Analysis}} (\bibinfo{publisher}{Cambridge
  University Press}, \bibinfo{year}{1985}{\natexlab{a}}).

\bibitem[{\citenamefont{Wolf}(2006)}]{wolf:010404}
\bibinfo{author}{\bibfnamefont{M.~M.} \bibnamefont{Wolf}},
  \bibinfo{journal}{Physical Review Letters} \textbf{\bibinfo{volume}{96}},
  \bibinfo{eid}{010404} (pages~\bibinfo{numpages}{4}) (\bibinfo{year}{2006}).

\bibitem[{\citenamefont{Horn and Johnson}(1985{\natexlab{b}})}]{H&J2}
\bibinfo{author}{\bibfnamefont{R.}~\bibnamefont{Horn}} \bibnamefont{and}
  \bibinfo{author}{\bibfnamefont{C.}~\bibnamefont{Johnson}},
  \emph{\bibinfo{title}{Topics in Matrix Analysis}}
  (\bibinfo{publisher}{Cambridge University Press},
  \bibinfo{year}{1985}{\natexlab{b}}).

\end{thebibliography}
\end{document}